\documentclass[12pt]{article}
\usepackage[final]{graphics}
\usepackage{amsmath}
\usepackage{amsfonts,amsbsy}
\usepackage[psamsfonts]{amssymb}
\usepackage{graphicx}
\usepackage{color}

\def\empile#1\above#2{\mathrel{\mathop{\kern 0pt#1}\limits_{#2}}}

\newcommand{\non}{\nonumber\\}

\newcommand{\GeV}{\,\hbox{GeV}}

\newcommand{\sll}{\raise.15ex\hbox{$/$}\kern-.43em\hbox{$l$}}
\newcommand{\slepsilon}{\raise.15ex\hbox{$/$}\kern-.53em\hbox{$\epsilon$}}
\newcommand{\slvarepsilon}{\raise.15ex\hbox{$/$}\kern-.53em\hbox{$\varepsilon$}}
\newcommand{\slL}{\raise.15ex\hbox{$/$}\kern-.53em\hbox{$L$}}
\newcommand{\slP}{\raise.15ex\hbox{$/$}\kern-.53em\hbox{$P$}}
\newcommand{\slp}{\raise.1ex\hbox{$/$}\kern-.63em\hbox{$p$}}
\newcommand{\slq}{\raise.1ex\hbox{$/$}\kern-.53em\hbox{$q$}}
\newcommand{\slv}{\raise.1ex\hbox{$/$}\kern-.63em\hbox{$v$}}
\newcommand{\slR}{\raise.15ex\hbox{$/$}\kern-.53em\hbox{$R$}}
\newcommand{\slQ}{\raise.15ex\hbox{$/$}\kern-.53em\hbox{$Q$}}
\newcommand{\slK}{\raise.15ex\hbox{$/$}\kern-.53em\hbox{$K$}}
\newcommand{\slk}{\raise.15ex\hbox{$/$}\kern-.53em\hbox{$k$}}
\newcommand{\slSigma}{\raise.15ex\hbox{$/$}\kern-.53em\hbox{$\Sigma$}}
\newcommand{\slcalP}{\raise.15ex\hbox{$/$}\kern-.63em\hbox{$\cal P$}}
\newcommand{\slA}{\raise.15ex\hbox{$/$}\kern-.73em\hbox{$A$}}
\newcommand{\slbfA}{\raise.15ex\hbox{$/$}\kern-.73em\hbox{${\imb A}$}}
\newcommand{\slpartial}{\raise.15ex\hbox{$/$}\kern-.53em\hbox{$\partial$}}
\newcommand{\sla}{\raise.15ex\hbox{$/$}\kern-.53em\hbox{$a$}}
\newcommand{\slb}{\raise.15ex\hbox{$/$}\kern-.53em\hbox{$b$}}
\newcommand{\slc}{\raise.15ex\hbox{$/$}\kern-.53em\hbox{$c$}}
\newcommand{\slD}{\raise.15ex\hbox{$/$}\kern-.53em\hbox{$D$}}
\newcommand{\slC}{\raise.15ex\hbox{$/$}\kern-.53em\hbox{$C$}}

\def\p{{\boldsymbol p}}
\def\q{{\boldsymbol q}}
\def\l{{\boldsymbol l}}
\def\k{{\boldsymbol k}}

\def\x{{\boldsymbol x}}

\def\r{{\boldsymbol r}}

\def\b{{\boldsymbol b}}

\def\wt{\widetilde}

\begin{document}

\thispagestyle{empty}

\title {\bf \bf Heavy quark pair production in high
energy \\pA collisions: Open heavy flavors}

\author{Hirotsugu Fujii and Kazuhiro Watanabe}
\maketitle
\begin{center}
Institute of Physics, University of Tokyo,\\ 
Komaba 3-8-1, Tokyo 153-8902, Japan
\end{center}

\begin{abstract}
\noindent
We study open heavy flavor meson production in proton-nucleus (pA)
collisions at RHIC and LHC energies within the Color Glass Condensate
framework. 
We use the unintegrated gluon distribution at small Bjorken's $x$
in the proton obtained by solving the
Balitsky-Kovchegov equation with running coupling correction
and constrained by global fitting of HERA data.
We change the initial saturation scale of the gluon distribution
for the heavy nucleus. 
The gluon distribution with 
McLerran-Venugopalan model initial condition is also used
for comparison.
We present transverse momentum spectra of
single $D$ and $B$ productions in pA collisions,
and the so-called nuclear modification factor.
The azimuthal angle correlation of 
open heavy flavor meson pair is also computed to study the
modification due to the gluon saturation
in the heavy nucleus at the LHC.
\end{abstract}

\section{Introduction}


Heavy quark production in high-energy proton-nucleus (pA)
collisions at RHIC and the LHC 
provides us with a unique opportunity to 
investigate the so-called
{\it parton  saturation} phenomenon\cite{Gribov:1984tu,Mueller:1985wy} 
at small Bjorken's $x$ in the incoming nucleus.
The large charm quark mass allows perturbative 
calculation of the quark production from the gluons,
while high center-of-mass collision energy $\sqrt{s}$ 
makes the relevant $x$ of the gluons still small.
These low-$x$ gluons are abundantly generated from
the larger-$x$ partons in view of the $x$ evolution.  
Then the saturation momentum scale $Q_s^2(x)$ emerges
dynamically as a semi-hard scale
below which virtuality ($Q^2 < Q_s^2(x)$)
coherence and nonlinearity of the $x$ evolution
become important.
This dynamics of small-$x$ degrees of freedom in hadrons
is described with the Color Glass Condensate (CGC)
effective theory~\cite{Gelis:2010nm}.

The saturation scale $Q_{sA}^2(x)$ in a heavy nucleus of the atomic mass
number $A$ is enhanced by the larger valence color charges seen at
moderate value of $x=x_0$. Indeed, the empirical
formula~\cite{StastGK1,GelisPSS1} $Q_{sA}^2(x)= Q_{s0}^2
A^{1/3}(x_0/x)^{\lambda}$ with $Q_{s0}^2=0.2 \GeV^2$, $x_0=0.01$ and
$\lambda=0.3$ suggests that the saturation scale is already comparable to
the charm quark mass $m_c \sim 1.5\GeV$ with $A=200$ at
RHIC energy $\sqrt{s}=200$ GeV. 
Therefore quantitative analysis of particle production 
in pA collisions will be very crucial\cite{prediction-pPb}.


In the previous paper~\cite{Fujii:2013gxa} we studied 
J/$\psi$ and $\Upsilon(1S)$ productions 
in pA collisions within the CGC framework\cite{BlaizGV2,FujiiGV2},
in order to quantify the effects of gluon saturation 
in a heavy nucleus on the heavy quark pair production.
We extend here our study to the production of
open heavy flavor mesons in pA collisions and will evaluate
$D$ and $B$ production cross-sections
differential in transverse momentum at mid and
forward rapidities, and azimuthal correlations of the 
$D\bar D$ pair ($B \bar B$ pair as well).
In the CGC framework,
multiple scatterings and gluon merging dynamics are encoded
in the effective unintegrated gluon distribution (uGD) 
function of a heavy nucleus.
These effects will cause relative depletion of quark
production yields and azimuthal momentum imbalance between 
the produced quark and antiquark.
They will be more prominent in the momentum region
lower than $Q_s^2(x)$.

This study is significant also in the context of 
nucleus-nucleus (AA) collisions
as a benchmark for discriminating the initial nuclear effects
from the subsequent hot-medium effects on heavy meson production,
presuming that no hot medium is formed in pA collisions.
Heavy flavor production in AA collisions
measured at RHIC~\cite{PHENIX1,PHENIX2} and the LHC~\cite{ALICE1,ALICE2}
shows a strong suppression 
(compared to that in pp collisions with appropriate normalization),
similar in magnitude to that of light hadrons,
which is interpreted as a large energy loss of the heavy quark in
a hot
medium~\cite{Dokshitzer:2001zm,Djordjevic:2003zk,Akamatsu:2008ge,TGBGP,Rapp:2009my}.
Initial nuclear effects should be
accounted properly here for precise evaluation of the medium effects.


Although azimuthal angle correlation measurement for charmed meson pair is
inaccessible at RHIC so far due to limited statistics,
LHCb collaboration recently measured 
the angle correlation at forward rapidity 
in pp collisions~\cite{double-charm,Maciula-Szczurek}.
We expect that it will become also available in AA collisions
at the LHC.
In AA collisions, 
the interactions of the heavy quarks with the hot medium will distort
the angle correlation of the pair
and may generate a new correlation by collective
flow\cite{NAGW}.
For a precise evaluation, again, we need to take account of the
initial state effects.


We use the quark pair production formula obtained in
Ref.~\cite{BlaizGV2,FujiiGV2} at LO in the
strong coupling constant $\alpha_s$ and the color charge
density $\rho_p$ in the proton, 
but including all orders in the color charge density $\rho_A$ 
in the nucleus. In the large $N$ limit, we only need 
the three point function $\phi^{q\bar q,g}$ of the gluons
in the nuclear target. We assume that $\phi^{q\bar q,g}$ can be obtained
from the dipole amplitude $S$, which seems 
also valid in the large $N$.

It is now standard to use non-linear Balitsky-Kovchegov (BK)
equation~\cite{Balit1,Kovch1} for describing the $x$ dependence 
of the uGD in a hadron.
It is argued that the inclusion of running coupling corrections
to the BK equation (now called rcBK equation) is essential to 
phenomenology\cite{Balit3,AlbacK1,Albacete1}.
Indeed, the rcBK equation with an appropriate initial
condition can fit the HERA DIS data quite well
~\cite{Albacete:2009fh,Albacete:2010sy} and are successful 
in reproducing/predicting the data at hadron colliders 
quantitatively~\cite{albacete-marquet,ALbacete:2010ad,Fujii:2011fh,Albacete:2012xq}.
We use the numerical solution of the rcBK equation
to describe the $x$ dependence of the gluon distribution in the
nuclear target in the present work 
in the same way as in our previous paper\cite{Fujii:2013gxa}.


This paper is organized as follows.
In Sec.~2 we briefly introduce the expression for production
cross-section of open heavy flavor mesons in pA collisions
within the CGC framework,
and the unintegrated gluon distribution obtained by solving
the rcBK equation. 
Next in Sec.~3 we present numerical results 
for single open heavy flavor production at RHIC and LHC energies, 
and also azimuthal angle correlation between the pair
of heavy mesons. Summary is given in Sec.~4.

\section{Heavy flavor meson production from CGC}

Heavy quark pair production cross-section of 
a quark with transverse momentum $\q$ and rapidity $y_q$ and 
an anti-quark with $\p$ and $y_p$ 
is given to the leading order in $\alpha_s$ and $\rho_p$
but full orders in $\rho_A$ as~\cite{BlaizGV2,FujiiGV2} 
\begin{align}
\frac{d \sigma_{q \bar{q}}}{d^2\p_\perp d^2\q_\perp dy_p dy_q}
= & \;
\frac{\alpha_s^2 N}{8\pi^4 (N^2 -1)} \; 
\frac{1}{(2\pi)^2}
\int\limits_{\k_{2\perp},\k_\perp}
\frac{\Xi(\k_{1\perp},\k_{2\perp},\k_{\perp})}
{\k_{1\perp}^2 \k_{2\perp}^2}
\;
\phi_{_A,y_2}^{q\bar{q},g}(\k_{2\perp},\k_\perp)
\; 
\varphi_{p,y_1}(\k_{1\perp})
\; ,
\label{eq:cross-section-LN}
\end{align}
where 
\begin{align}
\Xi(\k_{1\perp},\k_{2\perp},\k_{\perp})
= &
{\rm tr}_{\rm d}
\Big[(\slq\!+\!m)T_{q\bar{q}}(\slp\!-\!m)
\gamma^0 T_{q\bar{q}}^{\dagger}\gamma^0\Big]
\non
&
+
{\rm tr}_{\rm d}
\Big[(\slq\!+\!m)T_{q\bar{q}}(\slp\!-\!m)
\gamma^0 T_{g}^{\dagger}\gamma^0 + {\rm h.c.}\Big]
\non
&
+{\rm tr}_{\rm d} 
\Big[(\slq\!+\!m)T_{g}(\slp\!-\!m)\gamma^0 T_{g}^{\dagger}\gamma^0\Big]
\label{eq:Xi}
\end{align}
represents the relevant hard matrix element squared,
and transverse momentum conservation
$\k_{1\perp}  + \k_{2\perp} =\p_\perp+\q_\perp$
the parton level should be understood.
Here index 1 (2) refers to the quantity on the proton (nucleus) side. 
We have used the notation $\int_{\k_\perp} \equiv \int
d^2\k_\perp/(2\pi)^2$ for transverse momentum integration.
The $T_{q\bar{q}}$ term corresponds to the process where
the gluon from the proton splits into the quark-pair which then interacts
with the gluons in the target nucleus, while 
the $T_{g}$ term corresponds to the one where the gluon from the proton
interacts with the gluons in the nucleus and then splits into the quark pair.
The explicit expressions for $T_{q\bar{q}}$ and $T_{g}$ can be found 
in~\cite{BlaizGV2,FujiiGV2}.

The quark pair production cross-section 
(\ref{eq:cross-section-LN}) 
involves the gluon 2-, 3- and 4-point functions 
of the heavy nucleus in general\cite{BlaizGV2}, 
but in the large-$N$ limit we can express 
the cross-section with a single function
$\phi_{_A,y_2}^{q\bar{q},g}(\k_{2\perp},\k_\perp)$
in the form of Eq.~(\ref{eq:cross-section-LN}). 
Furthermore, we have assumed the translational invariance
in the transverse plane of the large nucleus, i.e., 
impact parameter dependence is simply ignored.
Within this approximation, the 3-point function
$\phi_{_A,y_2}^{q\bar{q},g}(\k_{2\perp},\k_\perp)$ 
describes the gluon
distribution in the nucleus and is expressed as\cite{BlaizGV2,FujiiGV2}
\begin{align}
\phi_{_A,_Y}^{q \bar{q},g}(\l_\perp,\k_\perp)
&=
\pi R_A^2 \; \frac{N\l^2_\perp}{4 \alpha_s} \; 
 S_{_Y}(\k_\perp) \;
S_{_Y}(\l_\perp-\k_\perp)\; ,  
\label{eq:3ptfn_2}
\end{align}
where $S_{_Y}(\k_\perp)$ is the Fourier-transformed  
dipole amplitude in the fundamental representation.
The uGD of the nucleus is obtained
by integrating over $\k_\perp$\cite{BlaizGV2}:
\begin{align}
\phi_{_A,_Y}^{g,g}(\l_\perp)=
\int_{\k_\perp} \phi_{_A,_Y}^{q\bar q,g}(\l_\perp,\k_\perp)
\; .
\end{align}
The  uGD $\varphi_{p,y}(\k_{1\perp})$ of the proton
is obtained by replacing the transverse area
$\pi R_A^2$ and the amplitude $S_Y$ with those for the proton.

Single quark production cross-section is obtained by integrating 
the pair production cross-section (\ref{eq:cross-section-LN})
over the anti-quark phase space:
\begin{align}
\frac{d \sigma_{q}}{d^2\q_\perp dy_q}
=
\int\frac{dp^+}{p^+} \; d^2 \p_\perp
\frac{d \sigma_{q \bar{q}}}{d^2\p_\perp d^2\q_\perp dy_p dy_q}
\; .
\label{eq:single-cross-section-LN}
\end{align}
Dividing the cross-section (\ref{eq:cross-section-LN}) or
(\ref{eq:single-cross-section-LN})
with the total inelastic 
cross-section $\sigma_{hadr}^{pA}$, which we estimate as
$\sigma_{hadr}^{pA} = \pi (R_A + R_p)^2 \approx\pi R_A^2$, 
we can obtain the average multiplicity per event\footnote{
The expression (\ref{eq:cross-section-LN}) is for {\it single}
quark-pair production. 
Ref.~\cite{double-charm} reports that
double charm production amounts to 10 \% of single charm production
in forward region in pp collisions at $\sqrt{s}=7$ TeV.}.
If we compute the multiplicity per event, 
the total inelastic cross-section is effectively 
cancelled out  with
the transverse size of nucleus $\pi R_A^2$ 
in $\phi_{_A,y_2}^{q\bar{q},g}(\k_{2\perp},\k_\perp)$
and therefore the proton size only remains explicitly 
in the expression of multiplicity. 
We set the proton size as $R_\text{p}=0.9$ fm
throughout this paper.

Energy dependence of the cross-section is implicit in the gluon
correlator $\phi_{_A,_Y}^{q\bar q,g}(\l_\perp,\k_\perp)$,
through the rapidity $Y=\ln(1/x)$ evolution of the
dipole amplitude 
\begin{align}
S_{_Y}(\x_\perp)\equiv
{\frac{1}{N}}{\rm tr}\big<{\wt U}(\x_\perp) 
{\wt U}^\dagger({\boldsymbol 0})\big>_{_Y}  
\; ,
\end{align}
where 
${\wt U}(\x_\perp)$  is the eikonal phase along the light-cone 
in the fundamental representation,
and $\langle\cdot\rangle_{_Y}$ indicates the average 
over the charge density distribution in the target at the scale $Y$.
Physically, $S_Y(\x)$ is the eikonal scattering matrix,
probed by a quark-antiquark pair moving along the light-cone direction
in the background gauge field in the target nucleus.
The amplitude $S_{_Y}(\x)$ obeys 
the BK equation\cite{Balit1,Kovch1}:
\begin{eqnarray}
&&-\frac{d}{dY}
S_{_Y}({\r_\perp})
 =
\int d\r_{1\perp} \,
\mathcal{K}(\r_\perp, \r_{1\perp}) \, 
\Big [
S_{_Y}({\r_\perp})
-
S_{_Y}({\r_{1\perp}})S_{_Y}({\r_{2\perp}})
\Big ]\; ,
\label{eq:Kovchegov}
\end{eqnarray}
where $\r_\perp = \r_{1\perp}+\r_{2\perp}$ and 
${\mathcal K}(\r_\perp,\r_{1 \perp})$ is the evolution kernel.
The BK equation is closed in the 2-point function $S_{_Y}(\x)$,
and therefore is numerically much easier to be handled.

It has been demonstrated~\cite{AlbacK1,Albacete1} that 
the BK equation with the running coupling corrections 
in Balitsky's prescription\cite{Balit3} but without
the subtraction term (rcBK equation):
\begin{align}
\mathcal{K}(\r_\perp,\r_{1\perp})=&
\frac{\alpha_s (r^2) N} {2\pi^2}\,
\left [
\frac{1}{r_1^2} \left ( \frac{\alpha_s(r_1^2)}{\alpha_s(r_2^2)}-1  \right )
+
\frac{r^2}{r_1^2 r_2^2}
+
\frac{1}{r_2^2} \left ( \frac{\alpha_s(r_2^2)}{\alpha_s(r_1^2)}-1  \right )
\right ]
\end{align}
includes the important part of the NLO corrections. 
The behavior of the resultant saturation scale is compatible with HERA data: 
$Q_s^2(Y)\propto \exp(\lambda Y)$ with $\lambda\approx 0.3$
\cite{StastGK1,GelisPSS1,Albacete1}.

Global fit analysis of the compiled HERA e+p data at $x<x_0=0.01$
was performed in \cite{Albacete:2009fh,Albacete:2010sy} 
using the rcBK equation  with
the initial condition at $x=x_0$ 
\begin{align}
S_{_{Y0}}(\r_\perp)=\exp \left [
-\frac{(r^2 Q_{s0,{\rm p}}^2)^\gamma }{4} \ln 
\left ( \frac{1}{\Lambda r} + e \right ) 
\right ] 
\; .
\end{align}
Here, in the evolution, 
we modify the infrared regularization of the 
running coupling in the coordinate space
to the smooth one\cite{Fujii:2011fh}:
\begin{align}
\alpha_s(r^2)= \left [b_0 \ln 
\left (\frac{4 C^2}{r^2 \Lambda^2}+a \right ) \right ]^{-1}
\end{align}
with $b_0=9/(4\pi)$.
The constant $a$ is introduced 
so as to freeze the coupling constant smoothly at
$\alpha_s(r \to \infty)=\alpha_{fr}$.
The parameter values are listed in Table~\ref{tab:par}. 
We also list a parameter set with the McLerran-Venugopalan (MV) model
initial condition $\gamma=1$, for comparison.

\begin{table}[t]
\renewcommand\arraystretch{1.2}
\begin{center}
\begin{tabular}{|c|cccc|}
\hline
set & $Q_{s0,\rm p}^2/{\rm GeV}^2$ & $\gamma$ & $\alpha_{fr}$ & $C$ \\
\hline 
g1118  & 0.1597 & 1.118 & 1.0 & 2.47 \\
MV     & 0.2    & 1     & 0.5 & 1\\
\hline
\end{tabular}
\caption{Parameter values of the dipole amplitude. $\Lambda=0.241$ GeV is fixed.
\label{tab:par}}
\end{center}
\end{table}

For a heavy nucleus $A$,
the saturation scale at moderate values of $x$ will be enhanced 
by a factor of the nuclear thickness $T_A(\b)$.
As we limit our analysis to mean bias events in this paper, 
we assume a simpler relation
\begin{eqnarray}
Q_{s,A}^2(x_0) = A^{1/3} \, Q_{s,p}^2(x_0) \; . 
\label{eq:GBW2}
\end{eqnarray}
We shall allow the saturation scale of the nucleus with $A=200$
in the range $Q_{s,A}^2=$ (4 -- 6)$\times Q_{s,{\rm p}}^2$
at initial point $x_0=0.01$.
The 3-point function in the nucleus at $x<x_0$ 
can be obtained from the numerical solution of 
the rcBK equation via Eq.~(\ref{eq:3ptfn_2}).
For $x_0 \le x \le 1$, on the other hand, 
we apply the following phenomenological Ansatz~\cite{FujiiGV2}:
\begin{equation}
\phi_{_A,_Y}^{q\bar{q},g}(\l_\perp,\k_\perp)=\phi_{_A,_{Y_0}}^{q\bar{q},g}(\l_\perp,\k_\perp)
\left(\frac{1-x}{1-x_0}\right)^4 \left(\frac{x_0}{x}\right)^{0.15}\; ,
\label{eq:largex-ansatz}
\end{equation}
where $Y_0\equiv\ln(1/x_0)$.

In the present paper
we compute the heavy meson production in pA collsions.
Heavy flavor meson pair production cross-section 
can be written as
\begin{align}
\frac{d \sigma_{h\bar h}}{d^2\q_{h\perp}  d^2\q_{\bar h\perp} dy_q dy_p}
=
f_{q\to h} f_{\bar q \to \bar h}
\int\limits_{z_\text{1min},z_\text{2min}}^1
dz_1 dz_2\; 
\frac{D_q^h(z_1)}{z_1^2}\frac{D_{\bar q}^{\bar h}(z_2)}{z_2^2}
\frac{d \sigma_{q\bar q}}{d^2\q_\perp d^2\p_\perp dy_q  dy_p}
\label{eq:double-open-cross-section-LN}
\end{align}
Here $\q_{h\perp}$ ($\q_{\bar h\perp}$) and $y_q$ ($y_{\bar q}$) are 
respectively transverse momentum and rapidity 
of the produced meson $h$ ($\bar h$). 
The longitudinal momentum fraction $z_1$ ($z_2$) of 
the heavy meson fragmented from the heavy quark (anti-quark) 
is defined as $q_{h\perp}=z_1q_\perp$ ($q_{\bar h\perp}=z_2p_\perp$).
The lower limit $z_\text{min}$ is set by the momentum fraction of
the meson fragmented from the heavy quark with
the maximum $q_\perp$ allowed kinematically.
Here we assume that the meson and the quark have the
same rapidity, $y_q=y_h$ ($y_p=y_{\bar h}$).

For the fragmentation function $D(z)$,  we use 
the Kartvelishvili fragmentation function\cite{kartvelishvili}, 
\begin{align}
D_q^h(z)=(\alpha+1)(\alpha+2) z^{\alpha}(1-z)
\; .
\end{align}
The value of $\alpha$ is set to
$3.5$ $(13.5)$ for $D$ $(B)$~\cite{fragmentation_ex,fragmentation_ex_B}.
The factor $f_{q\to h}$
represents the transition rate of the heavy quark $q$ 
fragmenting into the heavy meson $h$.
Empirical values, $f_{c\to D^0}=0.565$, $f_{c\to D^{\ast+}}=0.224$, 
and $f_{b\to \bar B^0}=0.401$ are taken
from \cite{Dmeson-ratio,Bmeson-ratio}.
For the charge conjugate states, we assume
$D_q^h(z)=D_{\bar q}^{\bar h}(z)$ and 
$f_{\bar q\to \bar h}=f_{q\to h}$.

Similarly single heavy meson production cross-section 
is expressed in convolution form of quark production
cross-section~(\ref{eq:single-cross-section-LN})
and the fragmentation function $D_q^h(z)$,
\begin{eqnarray}
\frac{d \sigma_{h}}{d^2\q_{h\perp} dy}
= f_{q\to h}
\int\limits_{z_\text{min}}^1dz\frac{D_q^h(z)}{z^2}
\frac{d \sigma_{q}}{d^2\q_\perp dy}
\; .
\label{eq:open-cross-section-LN}
\end{eqnarray}
Again we set $q_{h\perp}=zq_\perp$ and $y_q=y_{h}=y$.

\begin{figure}[tbp]
\centering
\resizebox*{!}{5.5cm}{\includegraphics[angle=270]{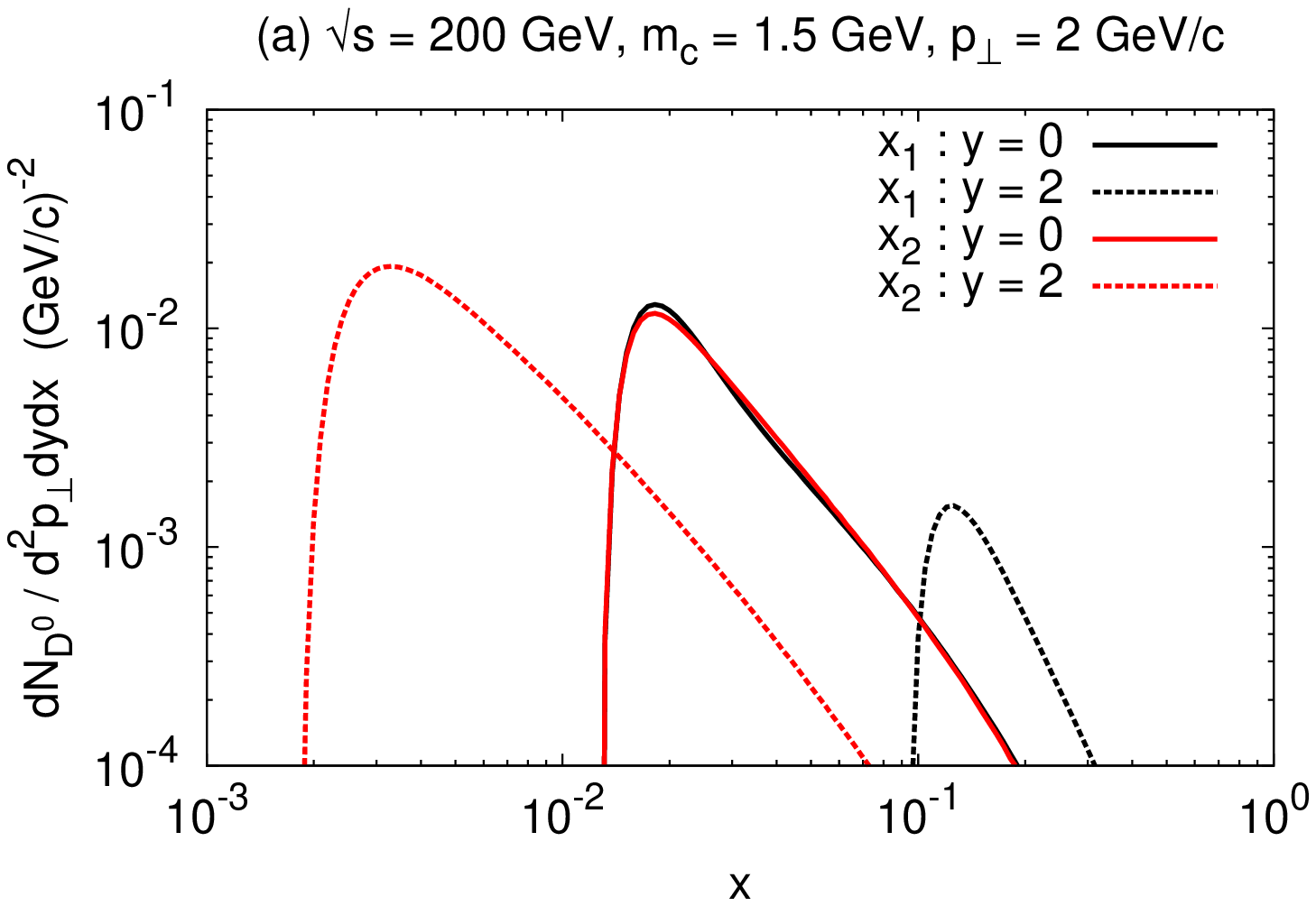}}
\resizebox*{!}{5.5cm}{\includegraphics[angle=270]{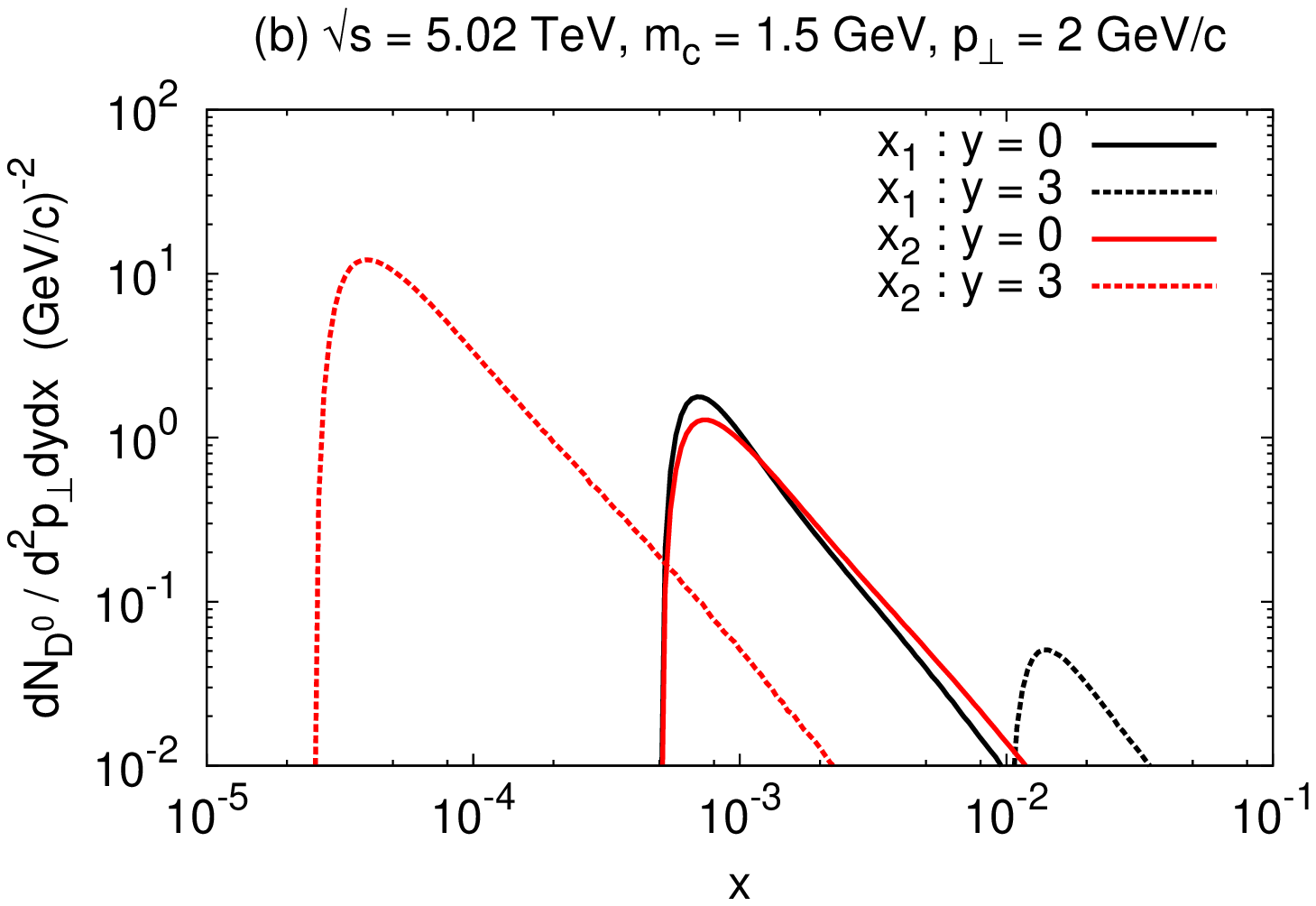}}
\resizebox*{!}{5.5cm}{\includegraphics[angle=270]{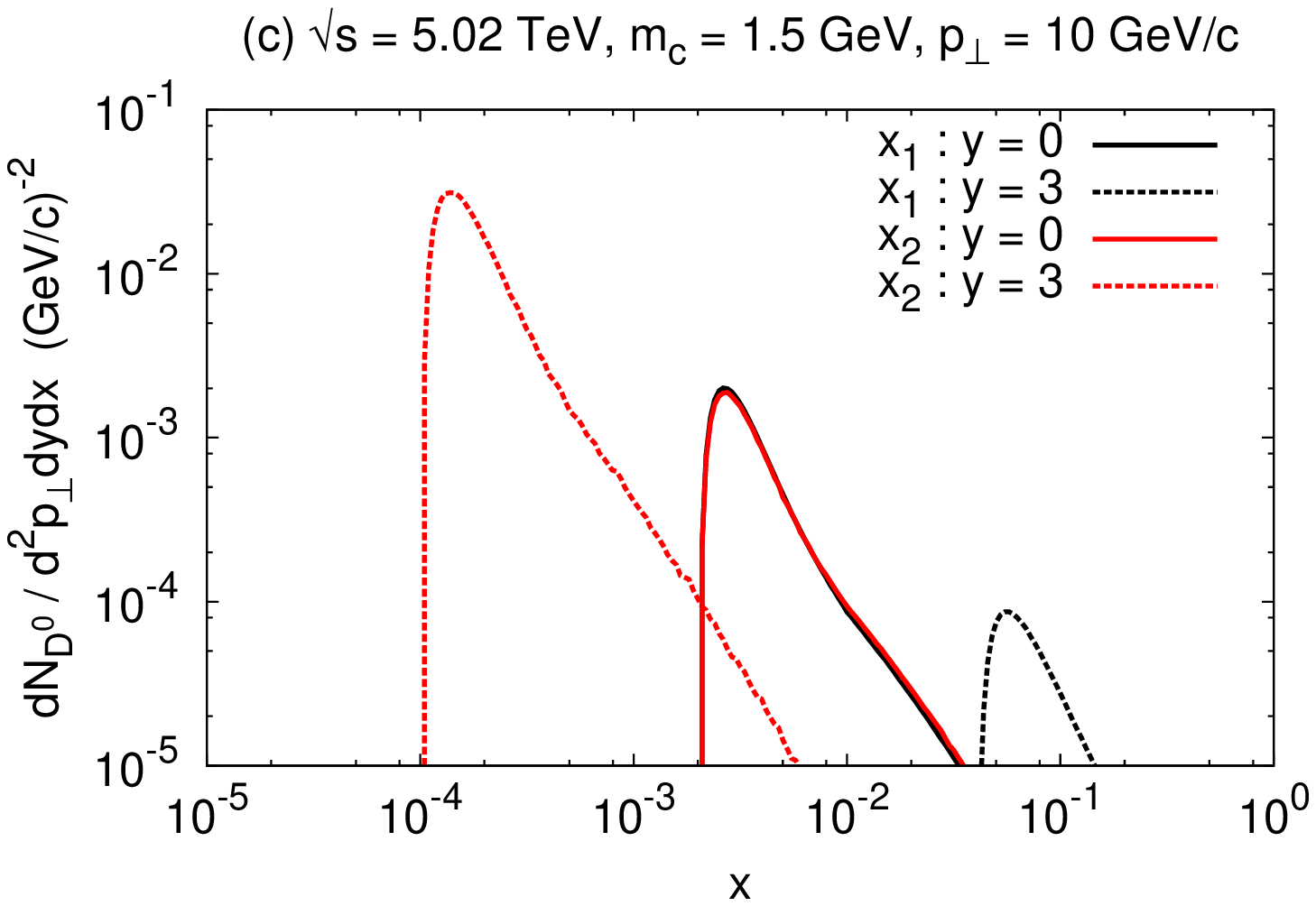}}
\resizebox*{!}{5.5cm}{\includegraphics[angle=270]{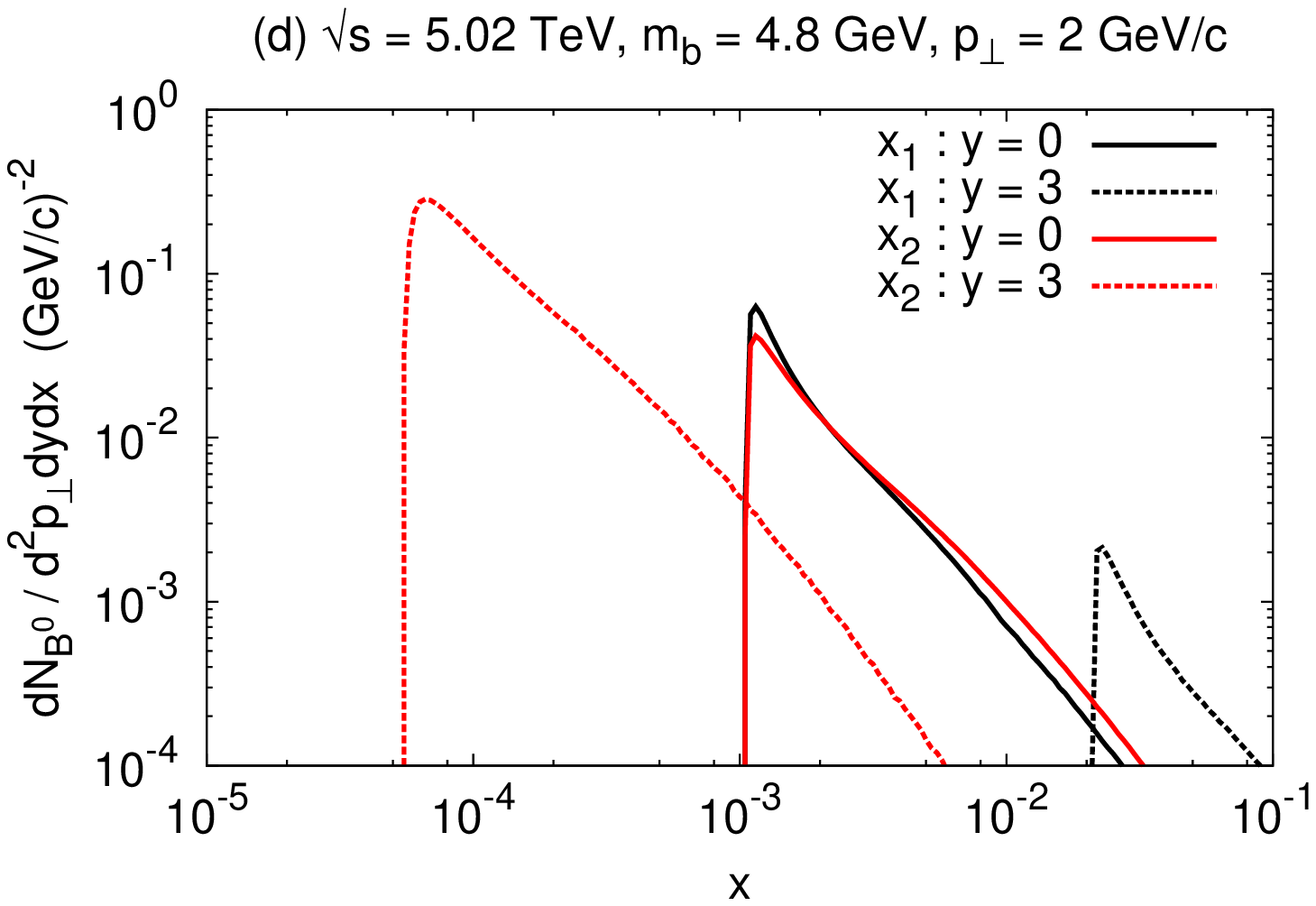}}
\caption{
$x_1$ (black) and $x_2$ (red)  coverages of $D^0$ production 
at mid and forward rapidities,
for fixed $p_\perp = 2$ GeV at $\sqrt{s}=200$ GeV (a),
and for fixed $p_\perp = 2$ (b) and 10 GeV (c) at $\sqrt{s}=5.02$ TeV.
$x_{1,2}$ coverages of $B^0$  production are shown in (d)
for fixed $p_\perp = 2$ at $\sqrt{s}=5.02$ TeV.
}
\label{fig:x2-coverage}
\end{figure}

Finally it would be instructive to show 
the kinematical coverage of $x$ variable in 
the heavy meson production at RHIC and LHC energies.  
We plot in Fig.~\ref{fig:x2-coverage} the 
$x_{1,2}$ distribution of single heavy meson production 
at a particular transverse momentum and rapidity.
We find in Fig.~\ref{fig:x2-coverage} (a) that both $x_1$ and $x_2$
contributing to single charmed meson production at $p_\perp=2$ GeV/c
and $y=0$ at $\sqrt s=200$ GeV are larger than $x_0=0.01$,
while at forward rapidity $y=2$ the production gets sensitivity to
small $x_2<x_0$.
In other words, the mid-rapidity production of single heavy mesons
is sensitive to the initial $\phi_{A,_Y0}^{\bar q q ,g}$ and
$x$-evolution effect shows up only at forward meson production
at RHICh energy.
However, it is seen
in Fig.~\ref{fig:x2-coverage} (b) and (c)
that at $\sqrt{s}=5.02$ TeV
small $x$ gluons around $10^{-3}$ dominate the production
even at mid rapidity.
In the forward-rapidity production, the $x_2$ value of the gluons from
the nucleus can become lower than $10^{-4}$, 
where one would expect good sensitivity of
heavy meson production to $x$-evolution and parton saturation.
Even for bottomed meson production
the situation is similar, as seen in Fig.~\ref{fig:x2-coverage} (d). 
Thus heavy quark productions, which may be evaluated with 
perturbation method, can be used to probe the small-$x$
dynamics by studying the heavy meson production at lower 
$p_\perp$  and forward rapidity at the LHC.

\newpage

\section{Numerical Results}

In numerical calculations,
we mainly use the uGD set g1118 in Table~\ref{tab:par}, 
and compare the results to
those with set MV and available experimental data.

\subsection{Transverse momentum spectrum}

\subsubsection{pp collisions}

\begin{figure}[tbp]
\begin{center}
\resizebox*{!}{5.5cm}{\includegraphics[angle=-90]{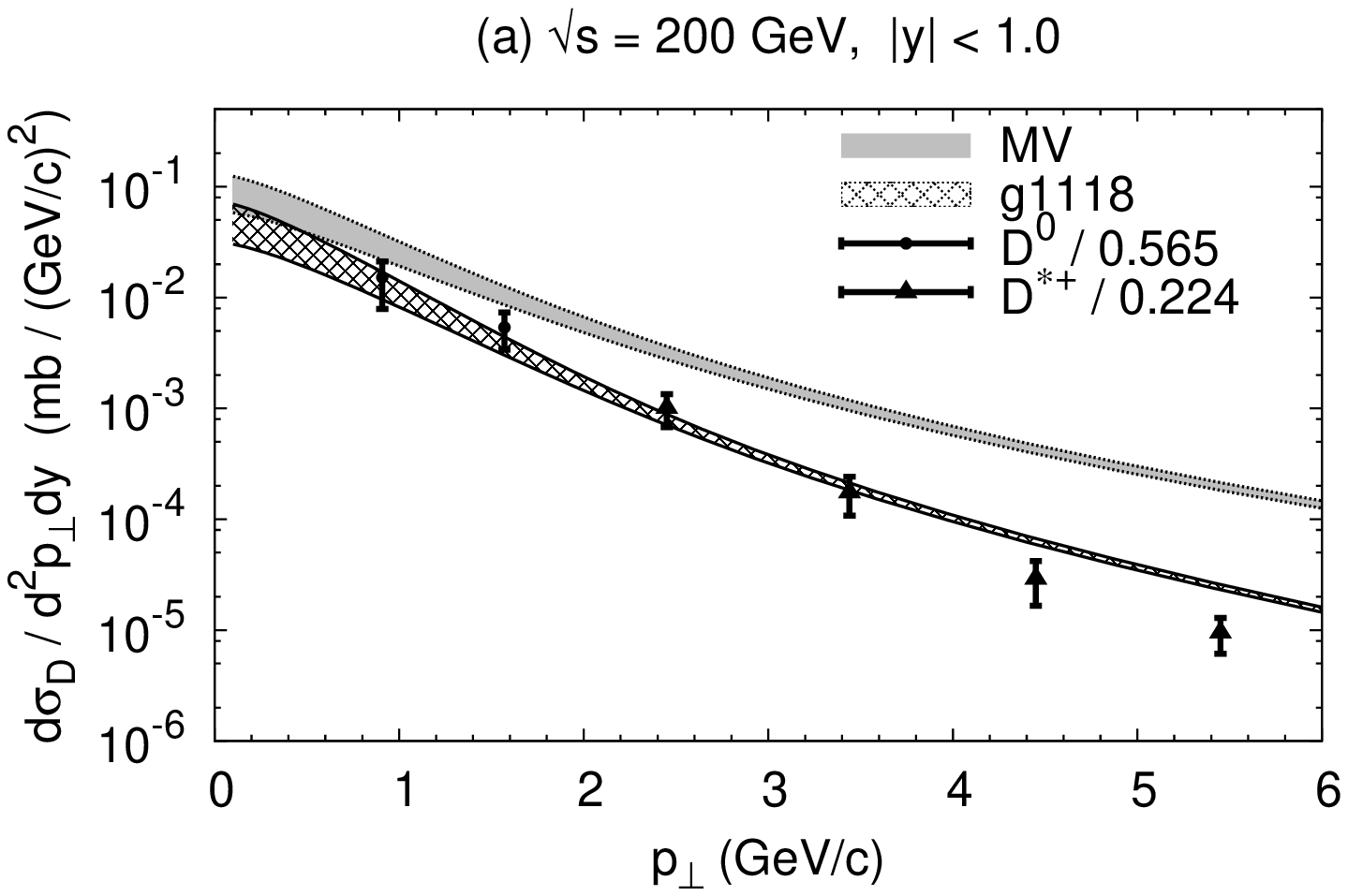}}
\resizebox*{!}{5.5cm}{\includegraphics[angle=-90]{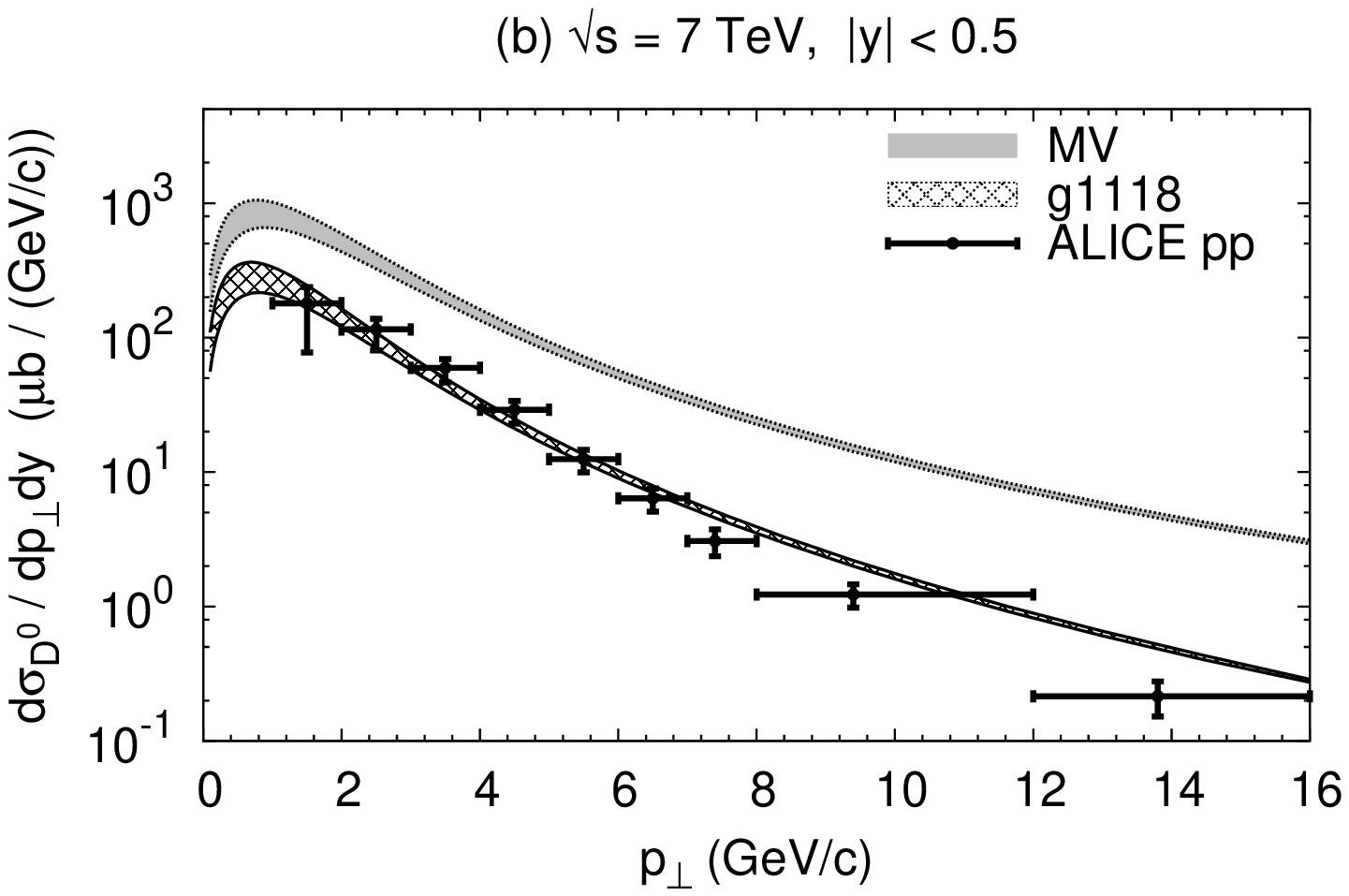}}
\end{center}
\caption{
(a) Differential cross-section of $D$ 
(rescaled as $D^0/f_{c\to D^0}$ and $D^{*+}/f_{c\to D^{*+}}$)
vs transverse momentum $p_\perp$
for rapidity range $|y|<1.0$ in pp collisions at $\sqrt{s}=200$ GeV, 
computed with Eq.~(\ref{eq:open-cross-section-LN})
with uGD sets MV (gray band) and g1118 (double-hatched).
The upper (lower) curve of the band corresponds to 
the result with $m_c=1.2$ (1.5) GeV.
The data is taken from Ref.~\cite{open_charm_xsection-STAR}.
(b) Differential cross-section of $D^0$ 
vs transverse momentum $p_\perp$
at $|y|<0.5$ in pp collisions at $\sqrt{s}=5.02$ TeV.
The ALICE data is taken from Ref.~\cite{ALICE:2011aa}.
}
\label{fig:D0-pp}
\end{figure}

\begin{figure}[tbp]
\begin{center}
\resizebox*{!}{5.5cm}
{\includegraphics[angle=-90]{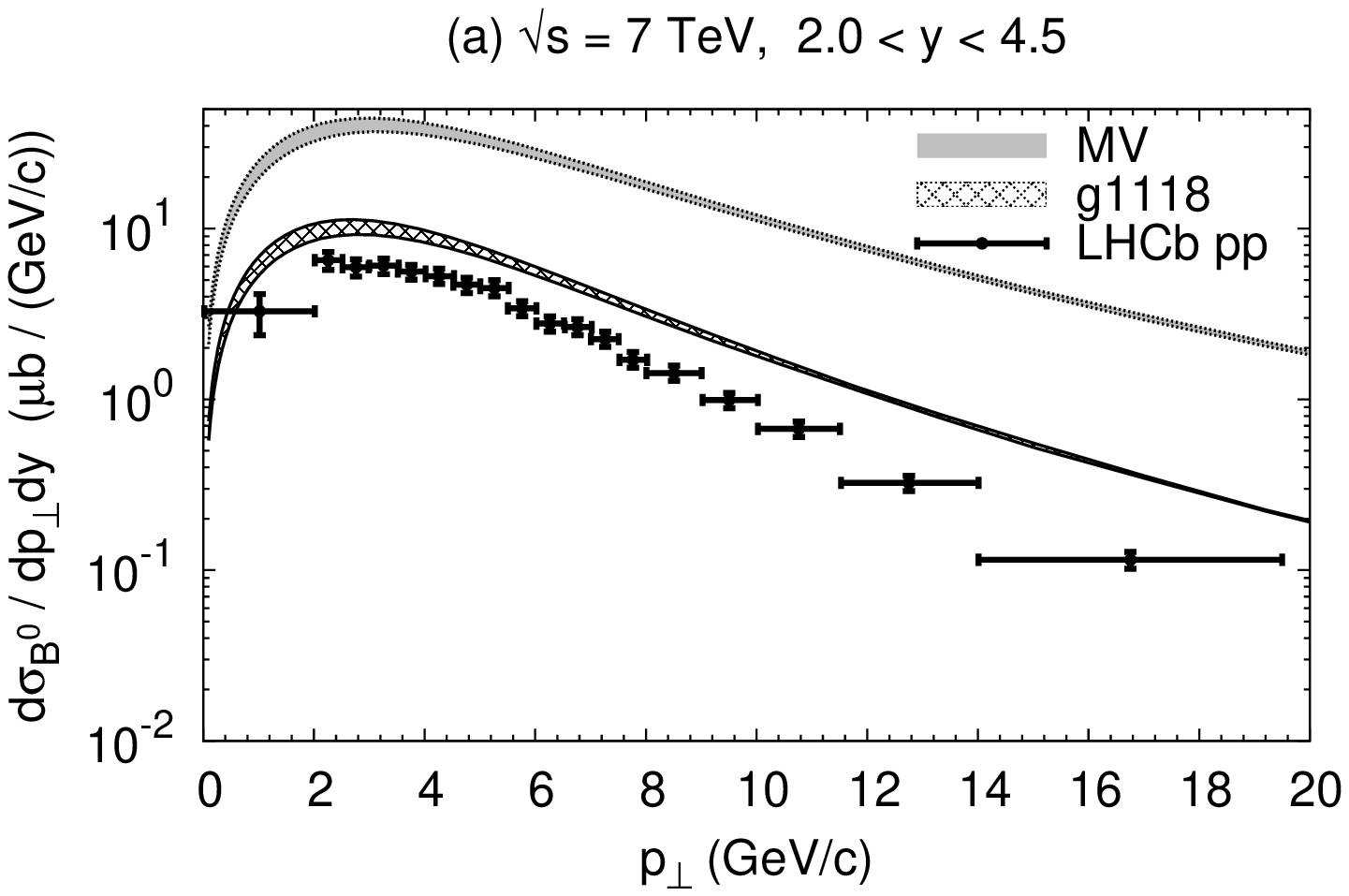}}
\resizebox*{!}{5.5cm}
{\includegraphics[angle=-90]{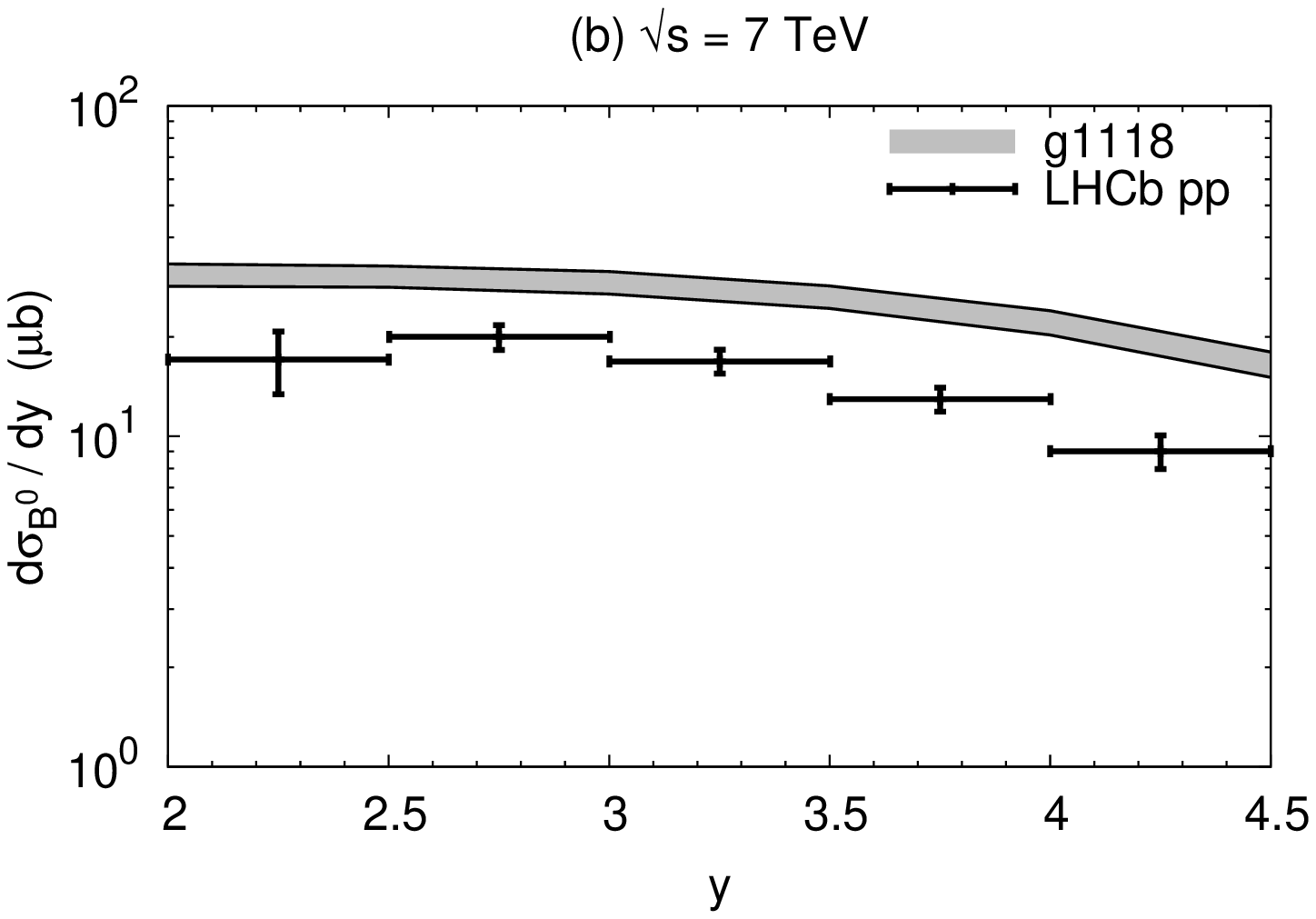}}
\end{center}
\caption{
(a) Differential cross-section of $B^0$ vs transverse momentum $p_\perp$
for rapidity range $2 < y < 4.5$ in pp collisions at $\sqrt{s}=5.02$ TeV,
computed with 
Eq.~(\ref{eq:open-cross-section-LN})
with uGD sets MV (gray band) and g1118 (double-hatched).
The upper (lower) curve of the band corresponds to 
the result with $m_b=4.5$ (4.8) GeV.
(b) Differential cross-section of $B^0$ vs $y$
in the range $0<p_\perp<40$ GeV
in pp collisions at $\sqrt{s}=5.02$ TeV.
The notation of the curve is the same as in (a).
The LHCb data is taken from Ref.~\cite{LHCb:B1}.
}
\label{fig:B0-pp}
\end{figure}

We study $D$ meson production cross-section
at mid rapidity in pp collisions at $\sqrt{s}=200$ GeV and 5.02 TeV.
Although the expression (\ref{eq:open-cross-section-LN}) is derived
for a dilute-dense system such as pA, 
we apply it here by substituting the numerical solution for the proton
into $\phi_{_A,_Y}^{q\bar q,g}(\l_\perp,\k_\perp)$.
By comparing the result with available data,
we can examine the applicability of our formula.
Furthermore we actually need the cross-sections in pp collisions
as the normalization 
when we study the nuclear modification of the cross-sections
in pA collisions.

We compute 
transverse momentum ($p_\perp$) spectrum of
$D$ meson production cross-section 
with uGD sets g1118 and MV in Table~\ref{tab:par}, 
and show the results in Fig.~\ref{fig:D0-pp}  together with 
the available data 
at $|y|<1$ and at $\sqrt{s}=200$ GeV\cite{open_charm_xsection-STAR}
and
at $|y|<0.5$ and at $\sqrt{s}=5.02$ TeV\cite{ALICE:2011aa}.
The upper (lower) curve of each band indicates the
result with charm quark mass $m_c=1.2$  (1.5) GeV.
We find that $p_\perp$ dependence of $D$ production
is better described with uGD set g1118, although it
gives still harder spectrum at high $p_\perp$.

Next we show forward $B^0$ production cross-section 
in $2 < y < 4.5$ in pp collisions at $\sqrt{s}=5.02$ TeV
as a function of $p_\perp$ in Fig.~\ref{fig:B0-pp} (a) 
and
the $p_\perp$-integrated cross-section 
as a function of $y$ in Fig.~\ref{fig:B0-pp} (b).
The upper (lower) curve of each band indicates the
result with the bottom quark mass $m_b=4.5$  (4.8) GeV.
The result with uGD set g1118 describes 
$p_\perp$ and $y$ dependences of the data~\cite{LHCb:B1}
better than that with set MV.
But the magnitude of cross-section is larger than the data
by about a factor of 2 -- 3.
We comment here that 
large-$x_1$ gluons in the proton become relevant 
in $B^0$ production at forward rapidity and/or at high $p_\perp$. 
Therefore the numerical result is sensitive 
to the extrapolation Ansatz Eq.~(\ref{eq:largex-ansatz})
of the uGD for large $x$.

\subsubsection{pA collisions}

\begin{figure}[tbp]
\begin{center}
\resizebox*{!}{5.5cm}
{\includegraphics[angle=-90]{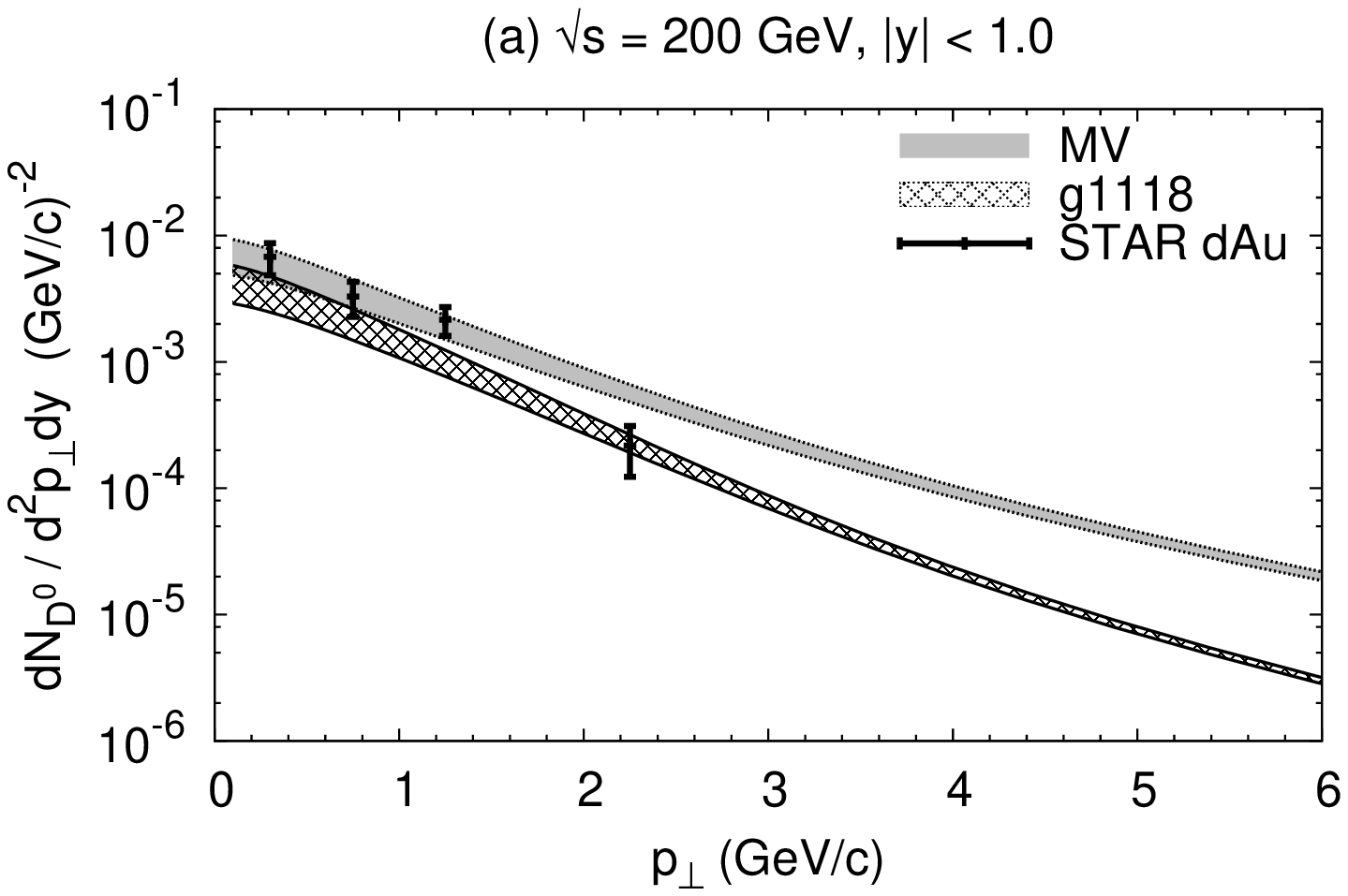}}
\resizebox*{!}{5.5cm}
{\includegraphics[angle=-90]{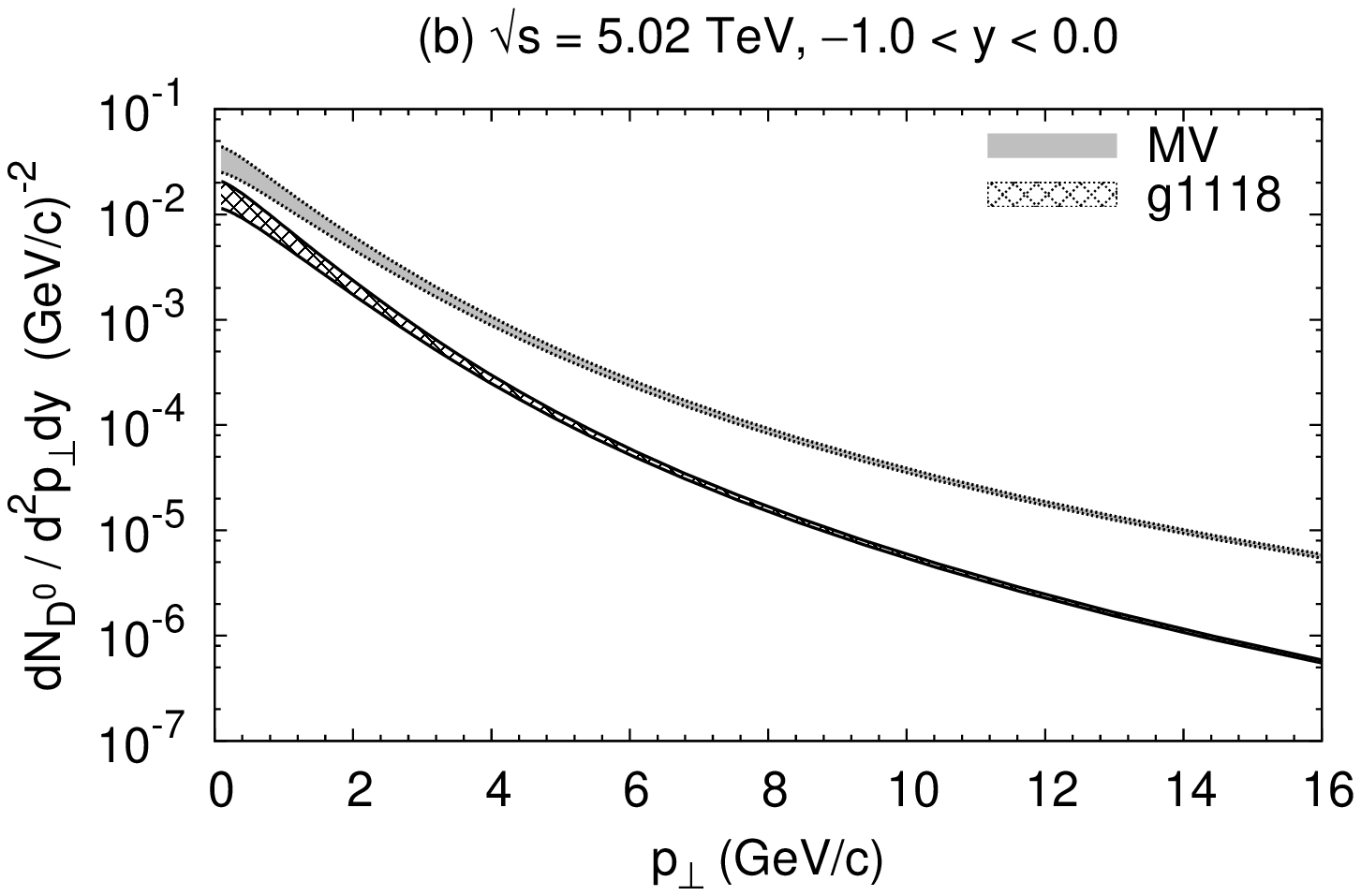}}
\end{center}
\caption{
Transverse momentum spectrum of $D^0$ multiplicity per event
in pA collisions, 
computed with Eq.~(\ref{eq:open-cross-section-LN})
with uGD sets MV (gray) and g1118 (double-hatch),
in rapidity range $|y|<1.0$  at $\sqrt{s}=200$ GeV (a)
and 
in $ -1.0<y<0.0$ at $\sqrt{s}=5.02$ GeV (b).
The upper (lower) curve of the band corresponds to 
the result with $m_c=1.2$ (1.5) GeV.
dAu data is taken from~\cite{open_charm_yields-STAR}.
}
\label{fig:D0-yield}
\end{figure}

We plot in Fig.~\ref{fig:D0-yield} (a) the 
transverse momentum spectrum of $D^0$ multiplicity
in the rapidity range $|y|<1.0$ in pA collisions 
at $\sqrt{s}=200$ GeV. 
We choose the initial saturation scale of the uGD 
in the heavy nucleus as $Q_{s0,A}^2(x=x_0)=6Q_{s0,p}^2$. 
The upper (lower) curve of the bands indicate the result
with $m_c=1.2$ (1.5) GeV.
We find that the results obtained with sets g1118 and MV 
fairly describe the available data 
at low $p_\perp\lesssim 2$ GeV~\cite{open_charm_yields-STAR}
although high-$p_\perp$ behaviors are different.
We show in Fig.~\ref{fig:D0-yield} (b) 
$D^0$ production spectrum in $-1<y<0$ at $\sqrt{s}=5.02$ TeV
\footnote{
Rapidity in the center-of-mass frame 
in pA collisions at $\sqrt{s}=5.02$ TeV
is shifted by $\Delta y=0.465$ from that in the laboratory frame.}.
The uGD sets MV and g1118 give different $p_\perp$ dependences
of the $D$ meson spectrum: Set MV yields harder $p_\perp$ spectrum.

\subsection{Transverse momentum dependence of $R_\text{pA}$} 

\begin{figure}[tbp]
\begin{center}
\resizebox*{!}{5.5cm}
{\includegraphics[angle=-90]{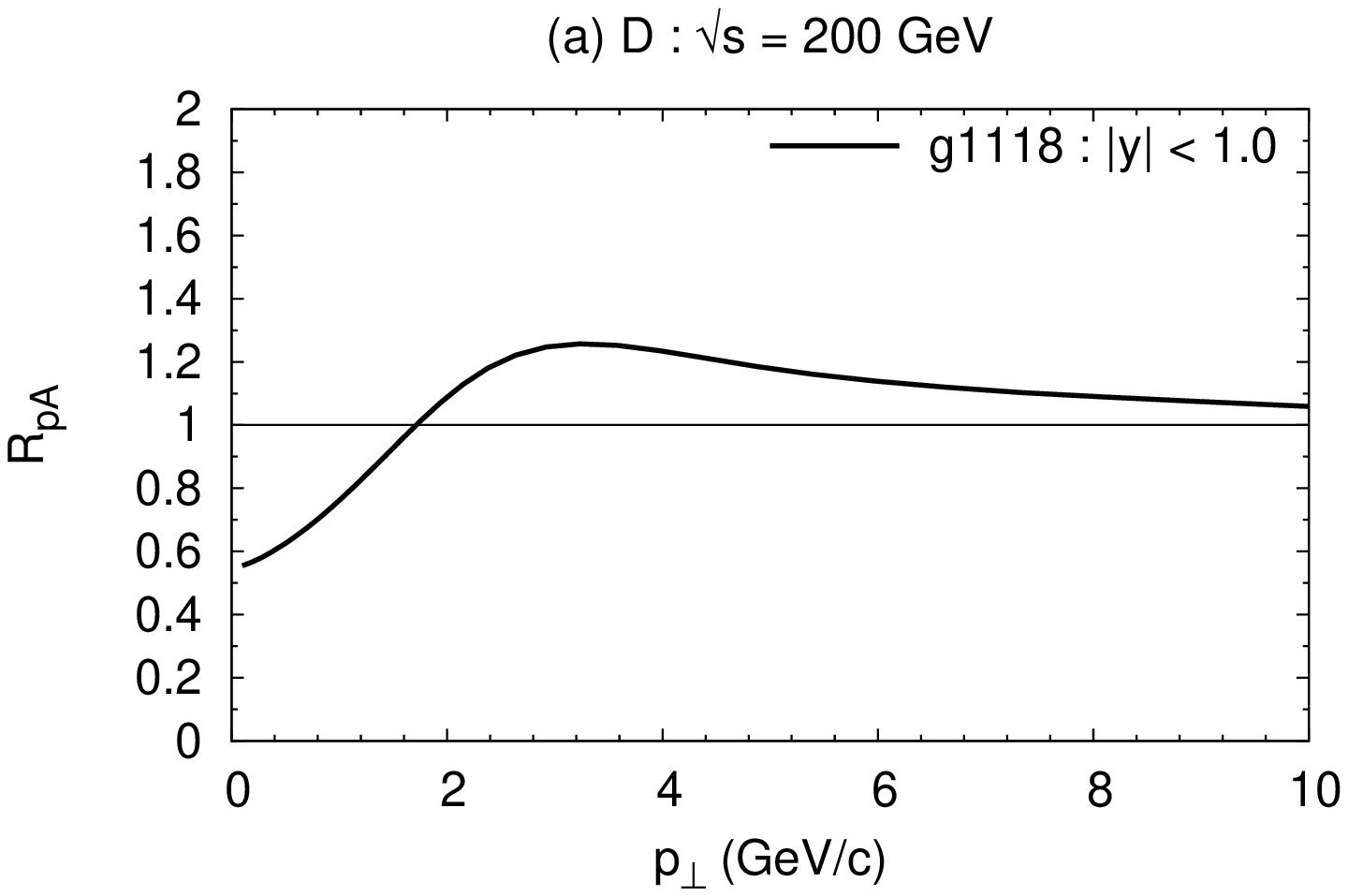}}
\resizebox*{!}{5.5cm}
{\includegraphics[angle=-90]{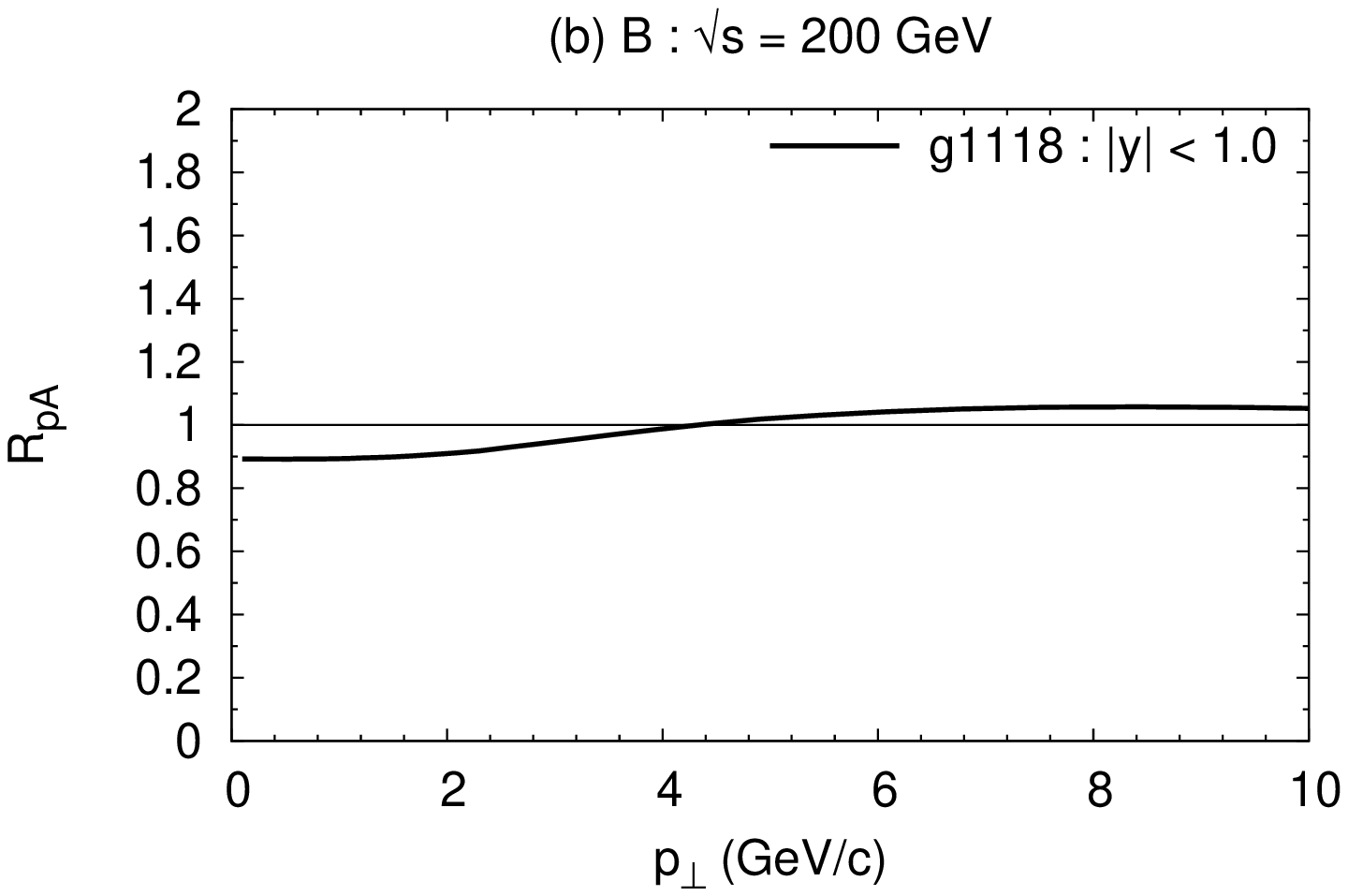}}
\end{center}
\caption{(a)~Nuclear modification factor $R_\text{pA}$  
of $D$ production vs $p_\perp$ 
computed with Eq.~(\ref{eq:def-RpA}) 
with uGD set g1118 with $m_c=1.5$ GeV
in the rapidity range $|y|<1.0$ at $\sqrt{s}=200$ GeV.
(b)~$R_\text{pA}(p_\perp)$ of $B$ production with $m_b=4.8$ GeV.
}
\label{fig:D-B-RpA-pt-RHIC}
\end{figure}

\begin{figure}[tbp]
\begin{center}
\resizebox*{!}{5.5cm}
{\includegraphics[angle=-90]{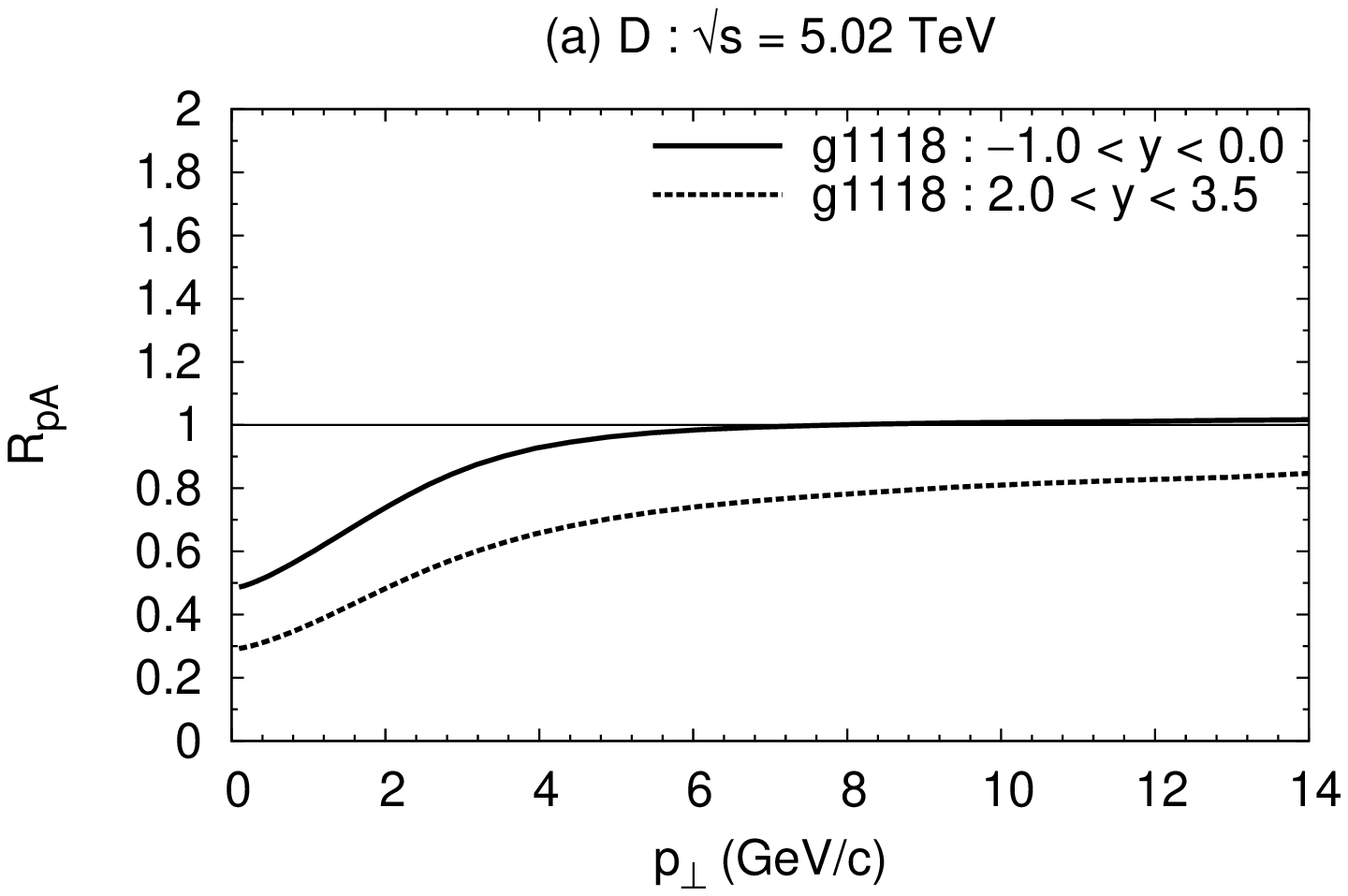}}
\resizebox*{!}{5.5cm}
{\includegraphics[angle=-90]{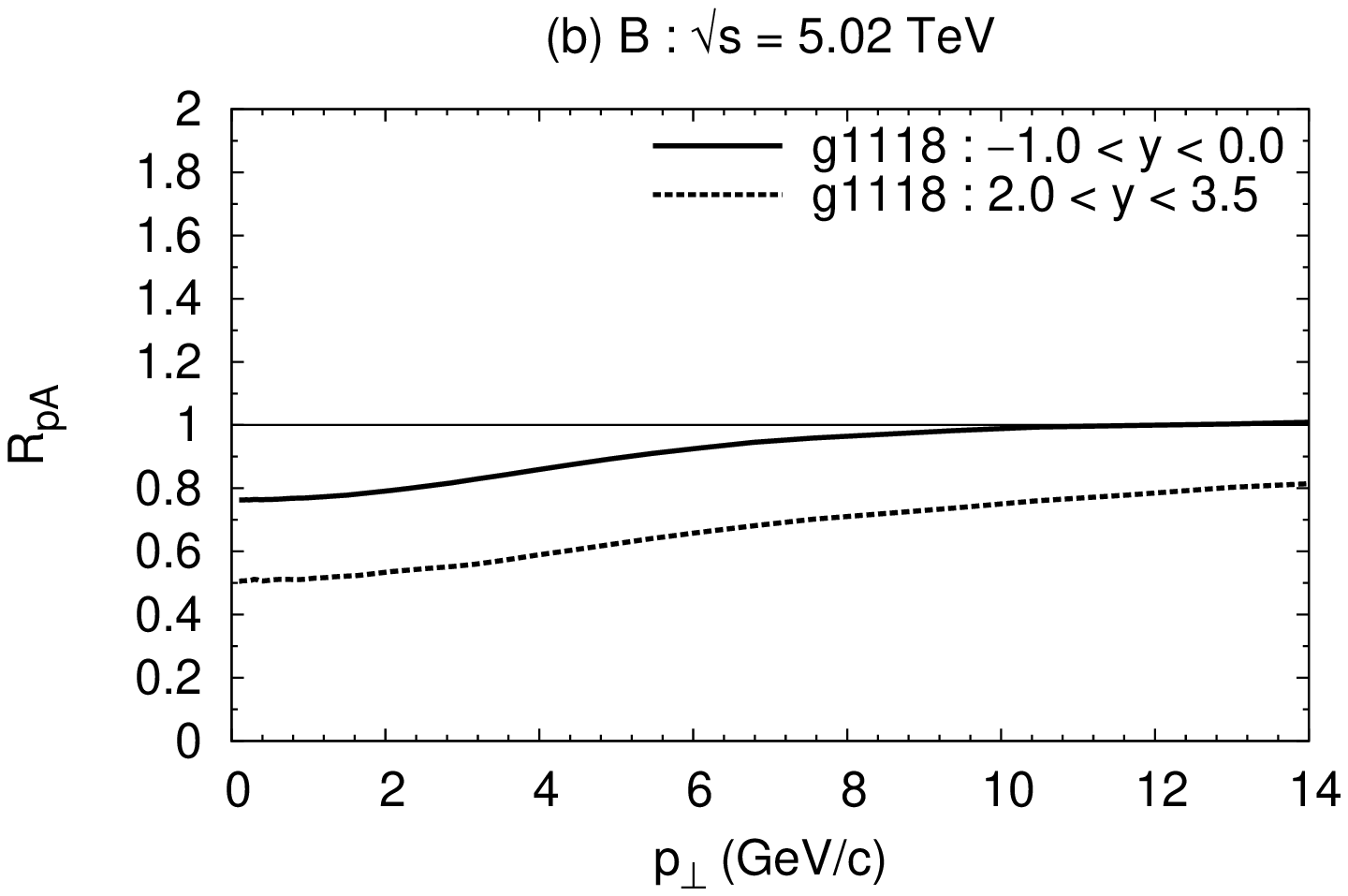}}
\end{center}
\caption{
Nuclear modification factor
$R_\text{pA}(p_\perp)$ of (a) $D$ and (b) $B$ productions 
for $-1 < y < 0$ (solid line) and $2<y<3.5$ (dotted line)
at $\sqrt{s}=5.02$ TeV.
}
\label{fig:D-B-RpA-pt-LHC}
\end{figure}

Now let us discuss the nuclear modification factor for pA collisions
defined as
\begin{align}
R_\text{pA}= \frac{\left . dN_{h}/d^2 p_\perp dy \right |_\text{pA}}
  {N_\text{coll}\, \left . dN_{h}/d^2 p_\perp dy \right |_\text{pp}}
\; .
\label{eq:def-RpA}
\end{align}
Here we set the number of binary nucleon-nucleon collisions
to $N_\text{coll}=A^{\gamma / 3}$\cite{Albacete:2012xq,Fujii:2013gxa}.
Model uncertainties in our calculation will largely cancel out
in the ratio of multiplicity per event in pA collisions 
to that in pp collisions.

In Fig.~\ref{fig:D-B-RpA-pt-RHIC} we plot
$R_\text{pA}$ of (a) $D$ and (b) $B$ productions
as a function of $p_{\perp}$ 
at mid rapidity ($|y|<1.0$) at $\sqrt{s}=200$ GeV.
We use the uGD set g1118 in this subsection.
We have checked that
$R_\text{pA}$ is insensitive to
the variation of the heavy quark mass within the range considered here,
and we show the results with $m_c=1.5$ GeV for $D$ production
and $m_b=4.8$ GeV for $B$ production.

The nuclear modification factor 
$R_\text{pA}$ of $D$ production is suppressed 
at lower $p_\perp \lesssim 2$ GeV 
while enhanced at higher $p_\perp \gtrsim 2$ GeV. 
As seen in Fig.~\ref{fig:x2-coverage} (a), heavy mesons are produced
from the gluons with moderate values of $x$, whose distribution 
is determined by the initial condition for $\phi_A$.
Thus the suppression and enhancement of $R_\text{pA}$ can be interpreted 
as the effects of the multiple scatterings in the nucleus encoded
in $\phi_A$. 
On the other hand,
$R_\text{pA}$ of $B$ production 
in Fig.~\ref{fig:D-B-RpA-pt-RHIC} (b) shows little structure as a function $p_\perp$.
This is because the larger bottom mass suppresses
the effects of multiple scatterings, i.e., $Q_{A,y0}^2 / m_b^2 \ll 1$.
That is, $B$ production scales with $N_\text{coll}$ at RHIC energy.

Next, we study
the nuclear modification $R_\text{pA}$ of $D$ and $B$ productions 
at $\sqrt{s}=5.02$ TeV.
$R_\text{pA}$ of $D$ production shown in Fig.~\ref{fig:D-B-RpA-pt-LHC} (a)
indicates that there is 
a strong suppression at lower $p_\perp$
and that no Cronin-like peak structure is seen at mid rapidity ($-1<y<0$)
by the quantum $x$-evolution effects on the small $x_2$ gluons.
We see the stronger suppression of $R_\text{pA}$ in the wider range of
$p_\perp$ at forward rapidity ($2<y<3.5$),
compared to that at mid rapidity.
At $\sqrt{s}=5.02$ TeV,
$B$ production at low $p_\perp$ shows a suppression similar to
but weaker than the $D$ production as shown in Fig.~\ref{fig:D-B-RpA-pt-LHC} (b).


\subsection{Rapidity dependence of $R_\text{pA}$}

\begin{figure}[tbp]
\begin{center}
\resizebox*{!}{5.5cm}
{\includegraphics[angle=-90]{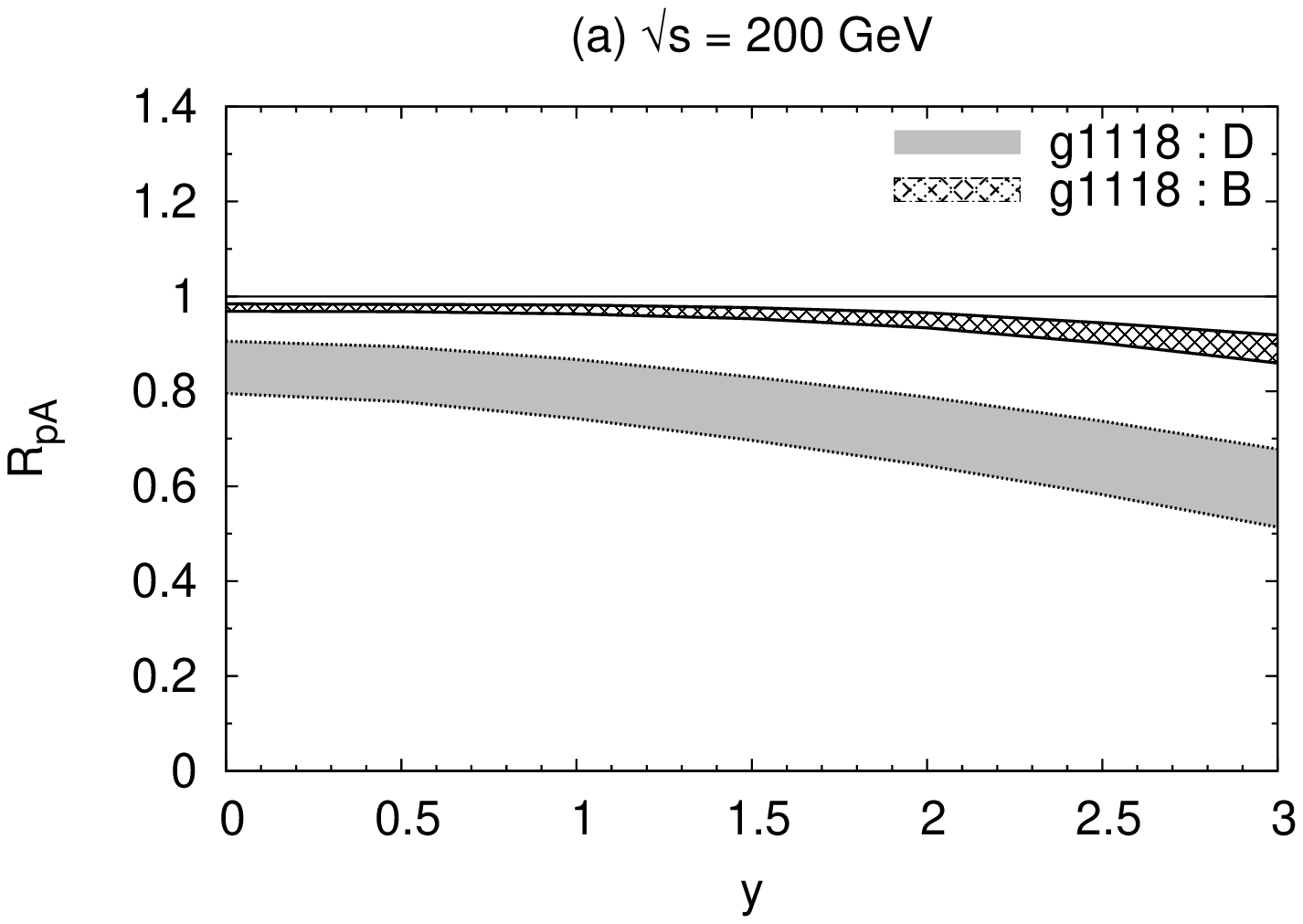}}
\resizebox*{!}{5.5cm}
{\includegraphics[angle=-90]{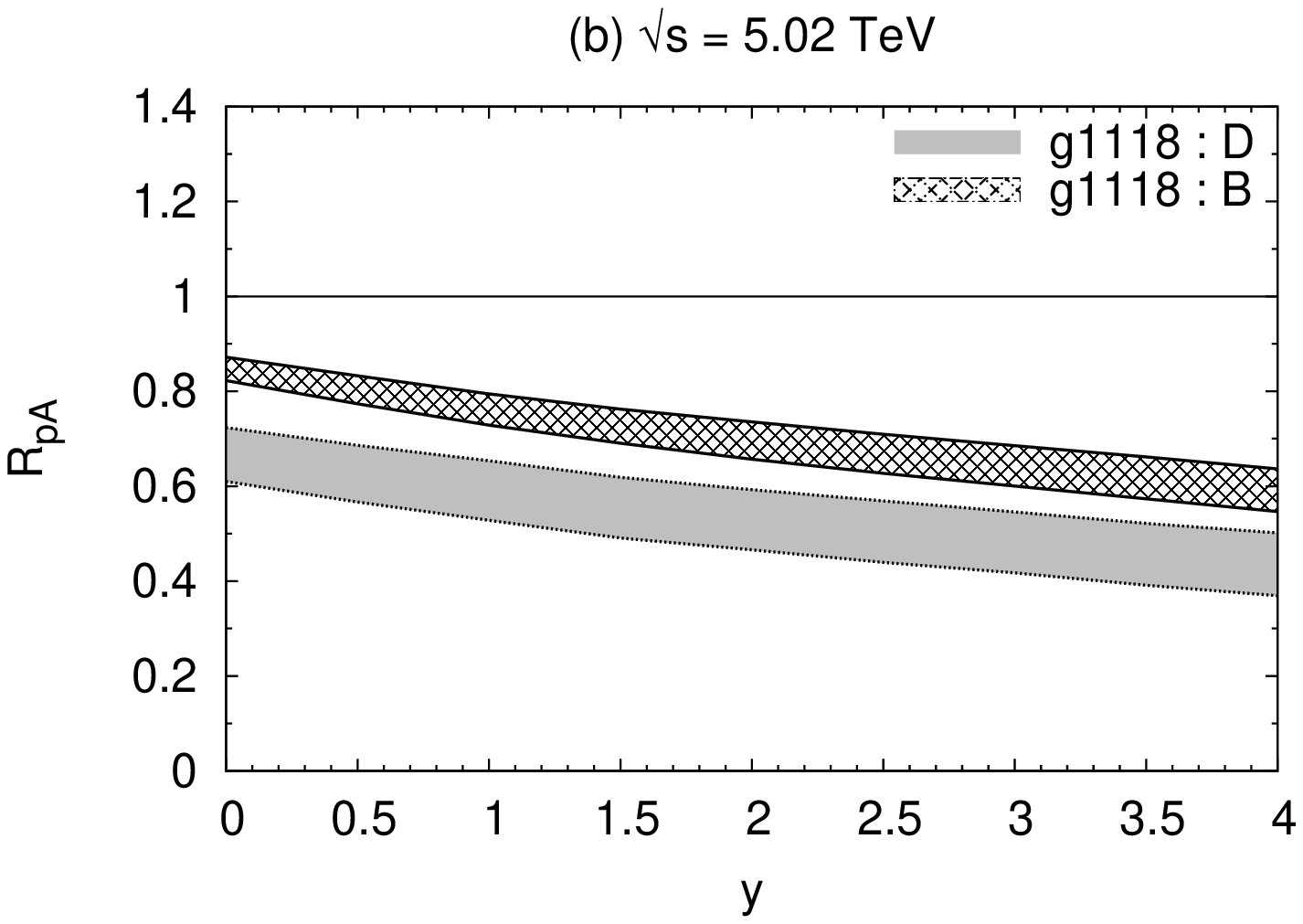}}
\end{center}
\caption{Nuclear modification factor $R_\text{pA}$ for $D$ (gray) and
$B$ (double-hatched) vs rapidity $y$ in pA collisions
at (a) $\sqrt{s}=200$ GeV and (b) $\sqrt{s}=5.02$ TeV.
The uGD set g1118 is used.
The bands indicate uncertainties from the variations 
$m_c=1.2-1.5$ GeV for $D$, and $m_b=4.5 - 4.8$ GeV for $B$
and also $Q_{s0,A}^2= (4-6) Q_{s0,p}^2$.
}
\label{fig:HQ-RpA-y}
\end{figure}

The nuclear modification factor ($R_\text{pA}(y)$)
of the heavy meson multiplicities $dN/dy$ in pA collisions
as a function of $y$ provides
important information about how the saturation effect evolves as 
moving to forward rapidity region.
In Fig.~\ref{fig:HQ-RpA-y} shown are the $R_\text{pA}$ of $D$ (gray band) and
$B$ (double hatched band) mesons
as a function of rapidity at  $\sqrt{s}=200$ GeV (a)
and 5.02 TeV (b).

We have allowed the variation of the initial saturation scale at $x=x_0$ 
in the heavy nucleus 
as $Q_{s0,A}^2(x=x_0)=(4-6) Q_{s0,p}^2$ with $A^{1/3}=4-6$ here.
The upper (lower) curve of the band of $D$ production
in Fig.~\ref{fig:HQ-RpA-y} now corresponds to the result 
with $m_c=1.5$ (1.2) GeV and $A^{1/3}=4$ (6).
For $B$ production, the upper (lower) curve corresponds to
the result obtained with $m_b=4.8 (4.5)$ GeV and $A^{1/3}=4 (6)$.
The width of the band here comes mainly from the change of $A^{1/3}=4-6$.

We find in Fig.~\ref{fig:HQ-RpA-y} (a) that $R_\text{pA}$ of  the $D$ production 
at mid rapidity at $\sqrt{s}=200$ GeV is suppressed, which reflects the multiple scattering effect 
as we have discussed in Fig.~\ref{fig:D-B-RpA-pt-RHIC}.
Stronger suppression of $D$ production is seen with increasing the rapidity, 
in accord with the quantum evolution of the gluon distribution $\phi_{A}$.
On the other hand, for $B$ production, 
$R_\text{pA}$ shows no appreciable change
with the increasing rapidity at RHIC energy,
besides a subtle suppression at very forward rapidities.

At $\sqrt{s}=5.02$ TeV, $R_\text{pA}$ of both $D$ and $B$ productions  
show large depletions even at mid rapidity as seen in Fig.~\ref{fig:HQ-RpA-y} (b).
Since the large colliding energy of the LHC
gives rise to much smaller $x_2<x_0$ of participating gluons
(Fig.~\ref{fig:x2-coverage} (b)--(d)), 
small-$x$ effects have already become relevant at mid rapidity,
and even $B$ production shows a suppression with increasing rapidity.


\begin{figure}[tbp]
\begin{center}
\resizebox*{!}{5.5cm}
{\includegraphics[angle=-90]{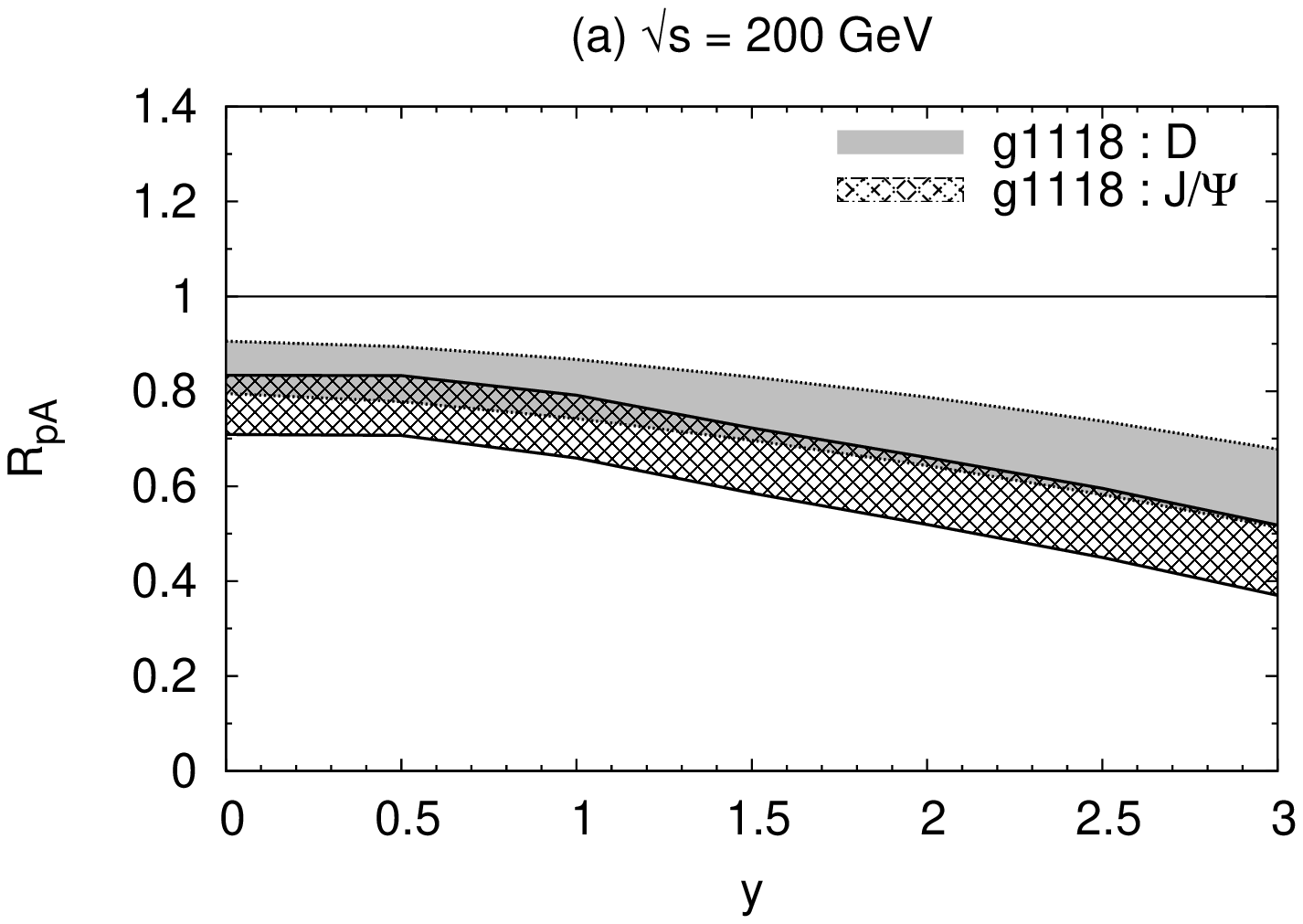}}
\resizebox*{!}{5.5cm}
{\includegraphics[angle=-90]{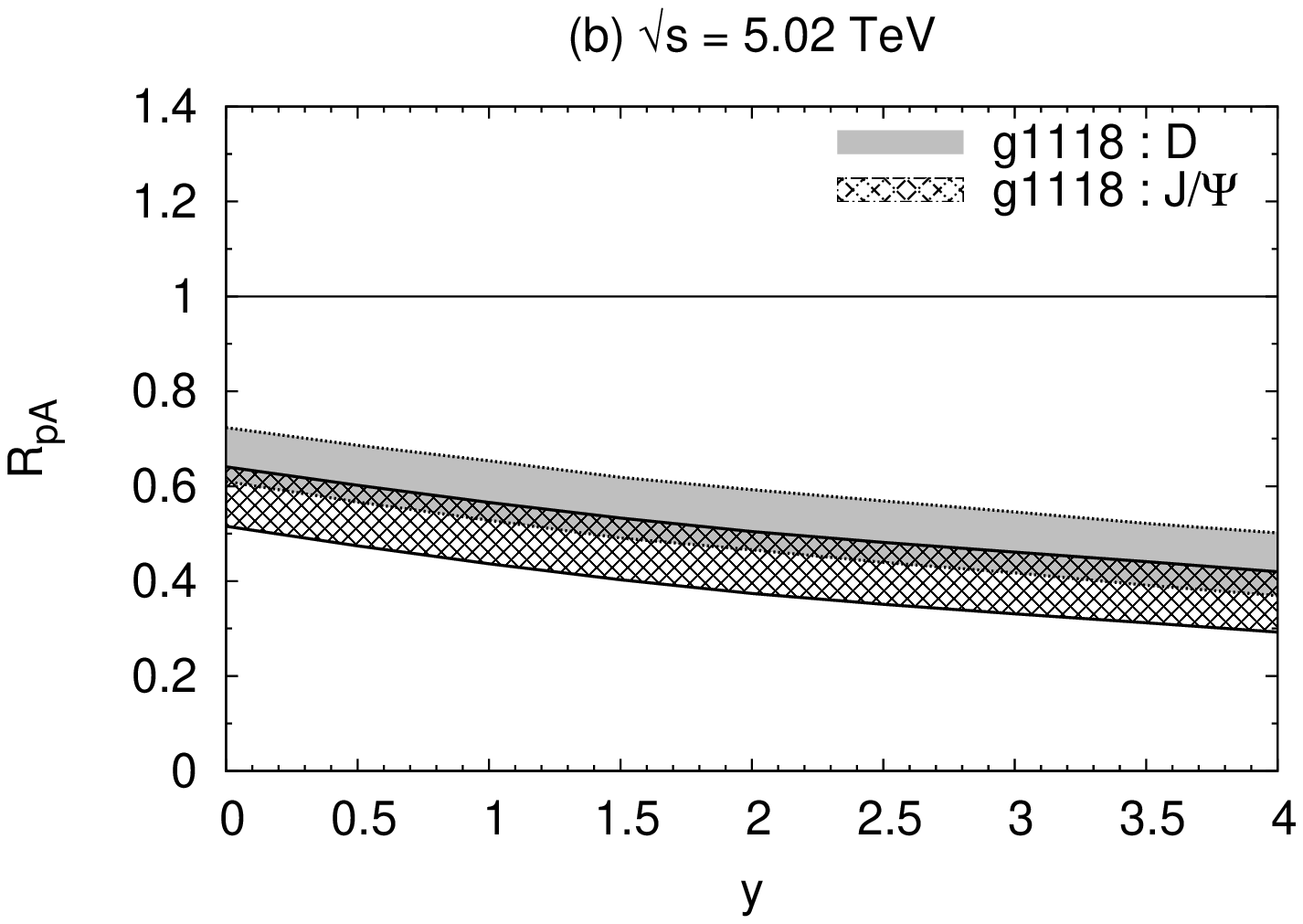}}
\end{center}
\caption{ Nuclear modification factor $R_\text{pA}$ for $D$ and J/$\psi$ 
vs $y$ in pA collisions at (a) $\sqrt{s}=200$ GeV
and (b) $\sqrt{s}=5.02$ TeV.
The bands indicate the uncertainties from the variations $m_c=1.2 - 1.5$ GeV 
and $Q_{s0,A}^2= (4-6) Q_{s0,p}^2$.
}
\label{fig:HQ-Quarkonium-RpA-y}
\end{figure}

Finally we compare  $R_{pA}$ for $D$ and J/$\psi$ productions 
as a function of rapidity at (a) $\sqrt{s}=200$ GeV and (b) $\sqrt{s}=5.02$ TeV
in Fig.~\ref{fig:HQ-RpA-y}.
We compute the J/$\psi$ production using color evaporation model, where
the heavy quark pair produced below the $D \bar D$ threshold is assumed to bound
into the quarkonium state with a fixed probability irrespective of the pair's color states.
(See Ref.~\cite{Fujii:2013gxa} for details).
We notice that J/$\psi$ production is more suppressed than $D$ meson.
This is because, in addition to the saturation effects of the initial gluons,
the produced quark pair experiences the multiple scatterings with the gluons in the target.
This effect increases the invariant mass of the pair on average.
In the color evaporation model, 
if the quark pair is kicked beyond the invariant mass threshold,
it cannot bound into the quarkonium, 
which results in a stronger suppression of 
the quarkonium than the $D$ meson production. 
We have also found that $\Upsilon(1S)$ is more suppressed than $B$ in our calculation, although it is not shown here.

\subsection{Azimuthal angle correlation}

Pair production of open heavy flavor covers wider kinematic
region of the participating partons than quarkonium production. 
In this subsection we examine nuclear modification 
of the azimuthal angle correlation of the heavy meson pair $h \bar h$
in pA collisions~\cite{albacete-marquet,dusling-venugopalan}.

We define the azimuthal angle correlation between $h$ and $\bar h$ as
the pair-production multiplicity per event
integrated over certain momentum and rapidity 
ranges with fixed angle $\Delta \Phi$ 
between the pair:
\begin{align}
CP[\Delta \Phi]
=
\frac{2\pi}{N_\text{tot}}\int p_{h\perp}dp_{h\perp} \,
 p_{\bar h\perp} dp_{\bar h\perp} dy_h dy_{\bar h}
\frac{dN_{h\bar h}}
     {d^2\p_{h\perp}d^2\p_{\bar h\perp}dy_{h}dy_{\bar h}}, 
\label{eq:correlation}
\end{align}
where $N_\text{tot}$ is the pair multiplicity per event integrated
over the same kinematic region 
and further integrated over the angle between the pair.
The pair production cross-section of the heavy mesons
is given in Eq.~(\ref{eq:double-open-cross-section-LN}).

\subsubsection{pp collisions}

\begin{figure}[tbp]
\begin{center}
\resizebox*{!}{5.5cm}{\includegraphics[angle=-90]{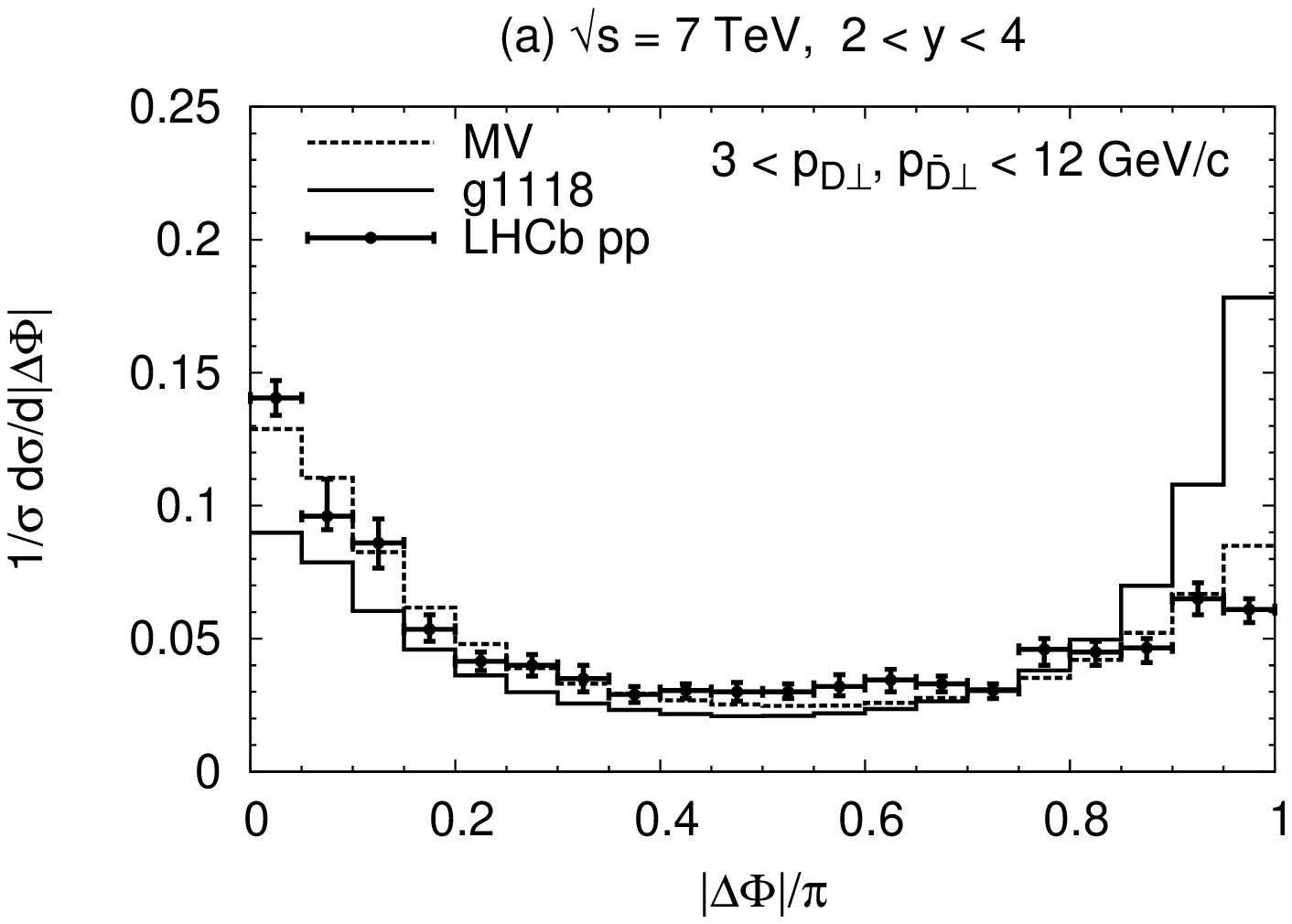}}
\resizebox*{!}{5.5cm}{\includegraphics[angle=-90]{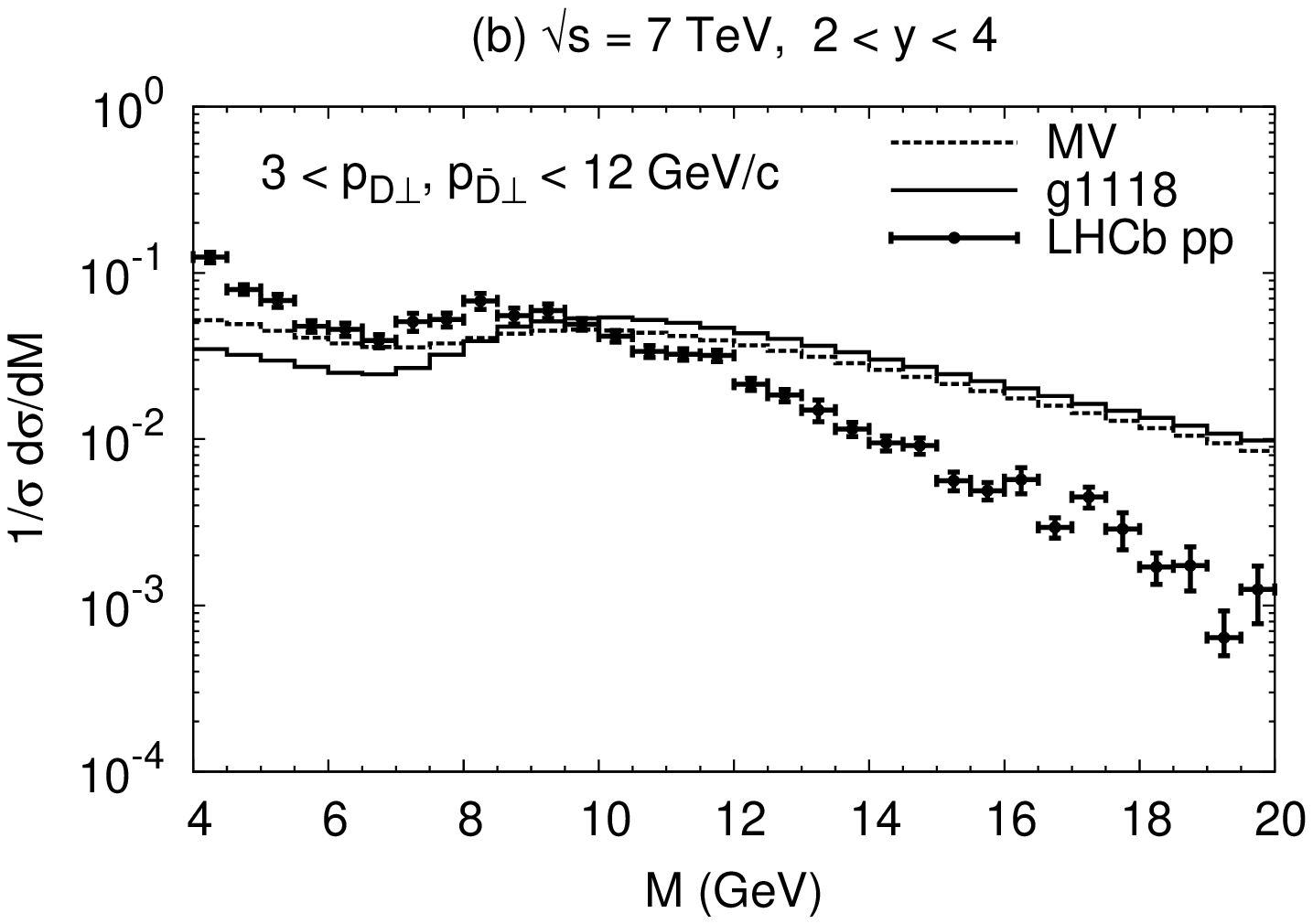}}
\end{center}
\caption{
Azimuthal angle correlation (a) and
invariant mass $M_{D\bar D}$ spectrum of $D^0\bar D^0$ pair production (b) 
in the rapidity and transverse momentum coverages, 
$2<y_D,y_{\bar D}<4$ and $3<p_{D\perp}, p_{\bar D\perp}<12 \GeV$
in pp collions at $\sqrt{s}=7$ TeV,
normalized by the total cross-section in the same fiducial region.
For binning, $\Delta \Phi/\pi=0.05$ in (a) and $\Delta M= 0.5$ GeV/$c^2$ in (b). 
Solid line denotes 
numerical result of Eq.~(\ref{eq:double-open-cross-section-LN}) with uGD set g1118,
the data points with error bars are taken from~\cite{double-charm}.
}
\label{fig:DDbar-LHCb}
\end{figure}

We compute the azimuthal angle correlation in  
$D^0\bar D^0$ pair production at the forward rapidity
in pp collisions at $\sqrt{s}=7$ TeV,
using the uGD sets g1118 and MV. 
We use $m_c=1.5$ GeV.
In order to compare the result with LHCb data~\cite{double-charm},
we set the kinematical range as $2<y_D, y_{\bar D}<4$ and 
$3<p_{D\perp}, p_{\bar D\perp}<12 \GeV$,
as plotted in Fig.~\ref{fig:DDbar-LHCb} (a). 
In \cite{double-charm} 
the bin size of the azimuthal angle 
is chosen as $\Delta \Phi/\pi =0.05$.

We immediately notice the near-side ($|\Delta \Phi|\sim 0$) and 
away-side ($|\Delta \Phi|\sim \pi$) enhancements in the numerical result.
The away-side peak is naturally expected 
from the back-to-back kinematics of the LO quark-pair production from two gluons
in the collinear factorization framework, but no near-side peak can be explained
unless the higher-order processes are considered.
In the CGC framework, on the other hand, gluon bremsstrahlung and multiple scatterings,
which are encoded in $\phi_{p}^{q\bar q,g}$,
provide {\it intrinsic transverse momentum} 
$k_\perp \sim Q_s$ of incident gluons.
This $k_\perp$ smears the away-side peak and 
generates the near-side peak in the angle correlation.
In the LHCb data, indeed, we see the near-side peak but an only subtle
away-side enhancement.
The numerical result with set MV fairly reproduces this
LHCb angle correlation, whereas in the result with set g1118 the away-side peak
still remains.
This is presumably reflecting the fact that 
the uGD set MV has harder $k_\perp$ spectrum than set g1118. 
But one should recall that the uGD set MV is already 
disfavored in the global 
fit~\cite{Albacete:2009fh,Albacete:2010sy}
and in hadron production analysis at collider 
energies~\cite{ALbacete:2010ad,Albacete:2012xq}.

The invariant mass spectrum of  $D^0\bar D^0$ pair production 
in pp collisions at $\sqrt{s}=7$ TeV is also measured 
in \cite{double-charm}.
We compare in Fig.~\ref{fig:DDbar-LHCb}~(b) our numerical results with the data.
The bin size for $M$ is 0.5 GeV.
The dip structure seen at low $M$ is understood as the effect of the lower momentum 
cut at 3 GeV/$c$.
Apparently the numerical result yields much harder invariant mass spectrum than
the observed data.

Several remarks are here in order:
First, large $M$ pair production probes the gluons at large $x_1$, 
where as explained in Sec. 2 we extrapolate the uGD with a simple 
Ansatz (\ref{eq:largex-ansatz}), which is likely to overestimate the uGD
in large $x$ region and needs more refinement.
Furthermore the back-to-back kinematics corresponds to the pair with
the large $M$ and small transverse momentum, where soft gluon emissions 
will be important and should be resummed~\cite{Mueller-Xiao-Yuan}.
Regarding small $M$ pair on the near-side, 
full NLO extension of the pair production formula 
may be important although
gluon splitting processes are partially included in the LO CGC formula
(\ref{eq:cross-section-LN}).

\subsubsection{pA collisions}

\begin{figure}[tbp]
\begin{center}
\resizebox*{!}{5.5cm}{\includegraphics[angle=-90]{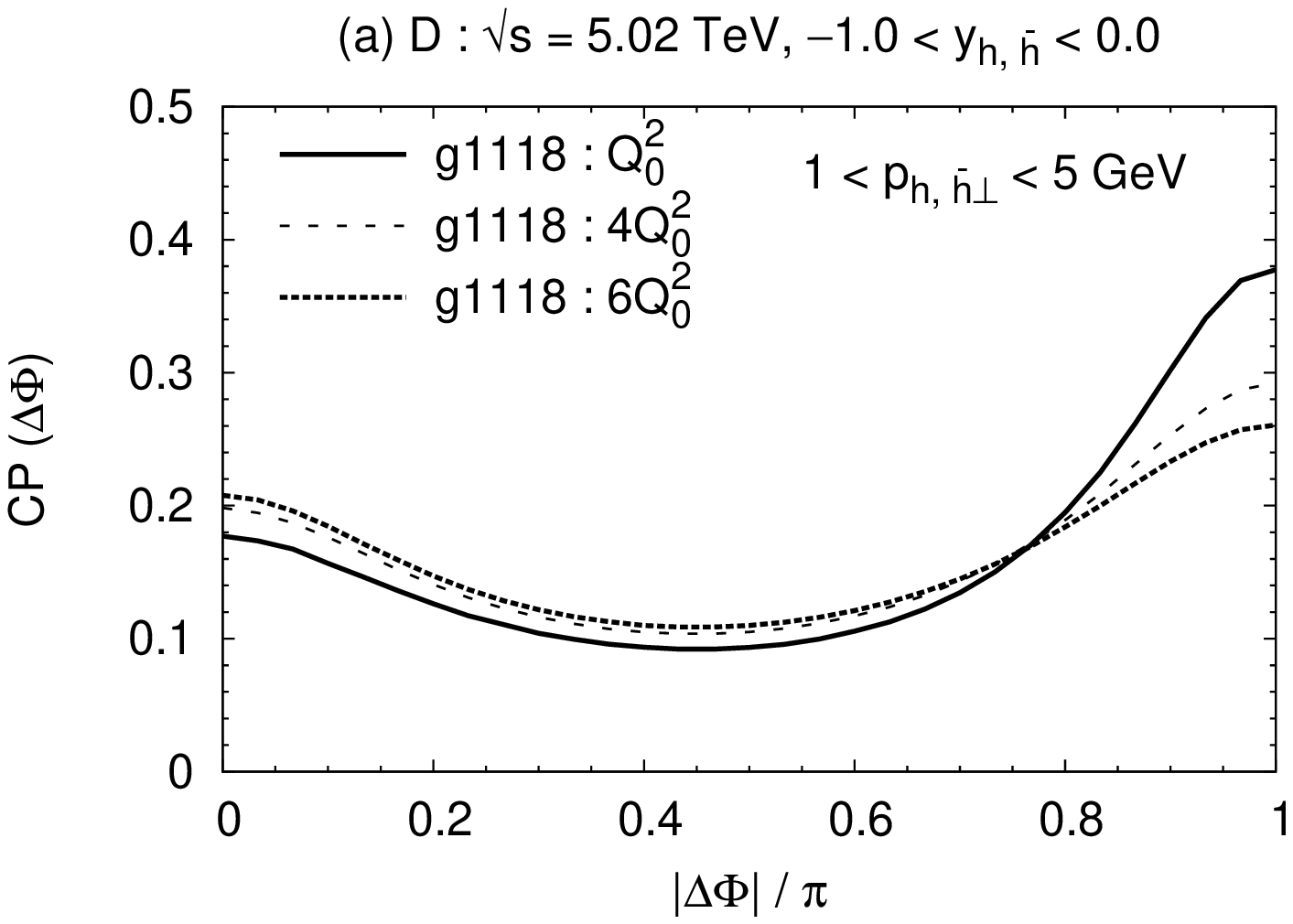}}
\resizebox*{!}{5.5cm}{\includegraphics[angle=-90]{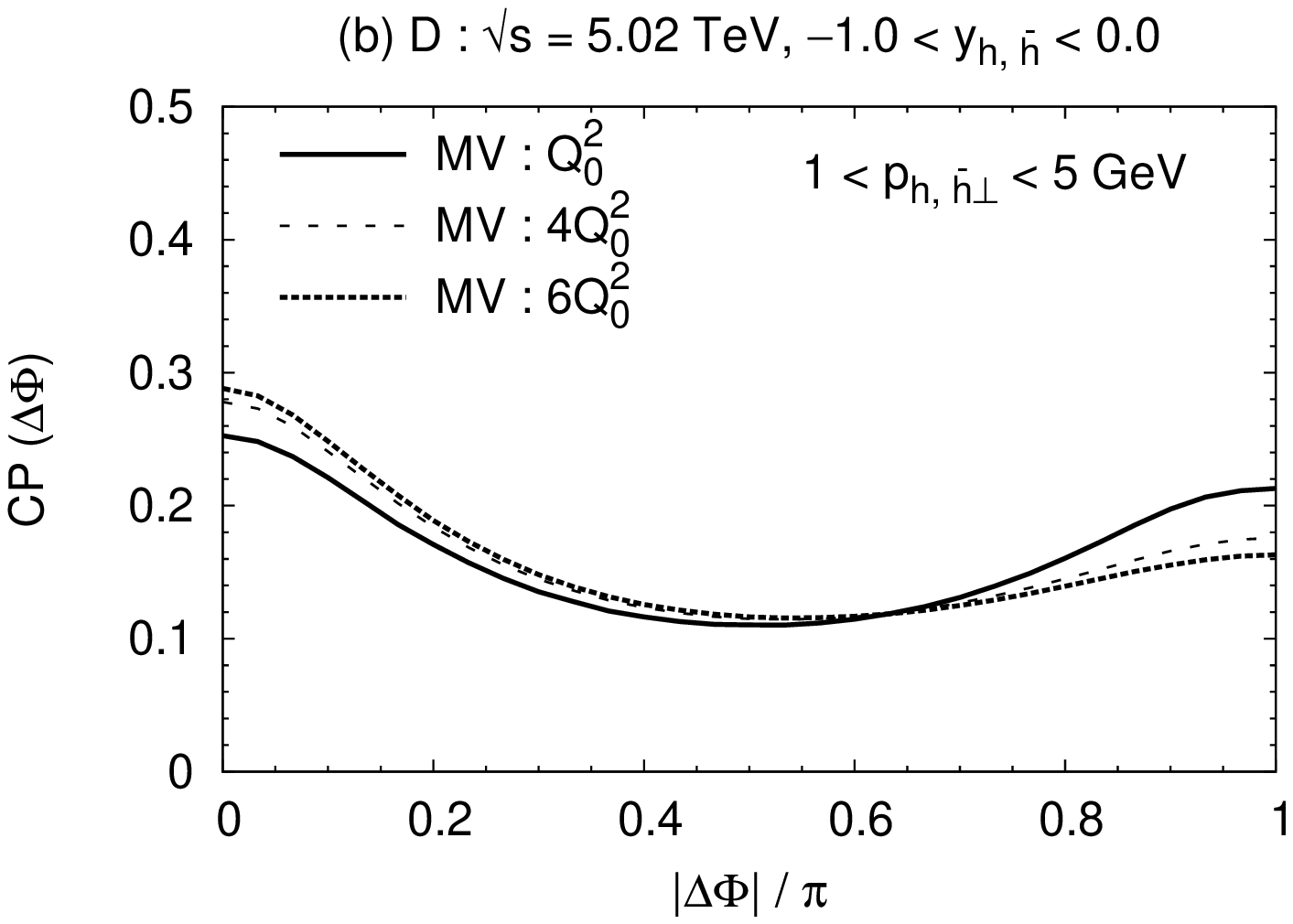}}
\resizebox*{!}{5.5cm}{\includegraphics[angle=-90]{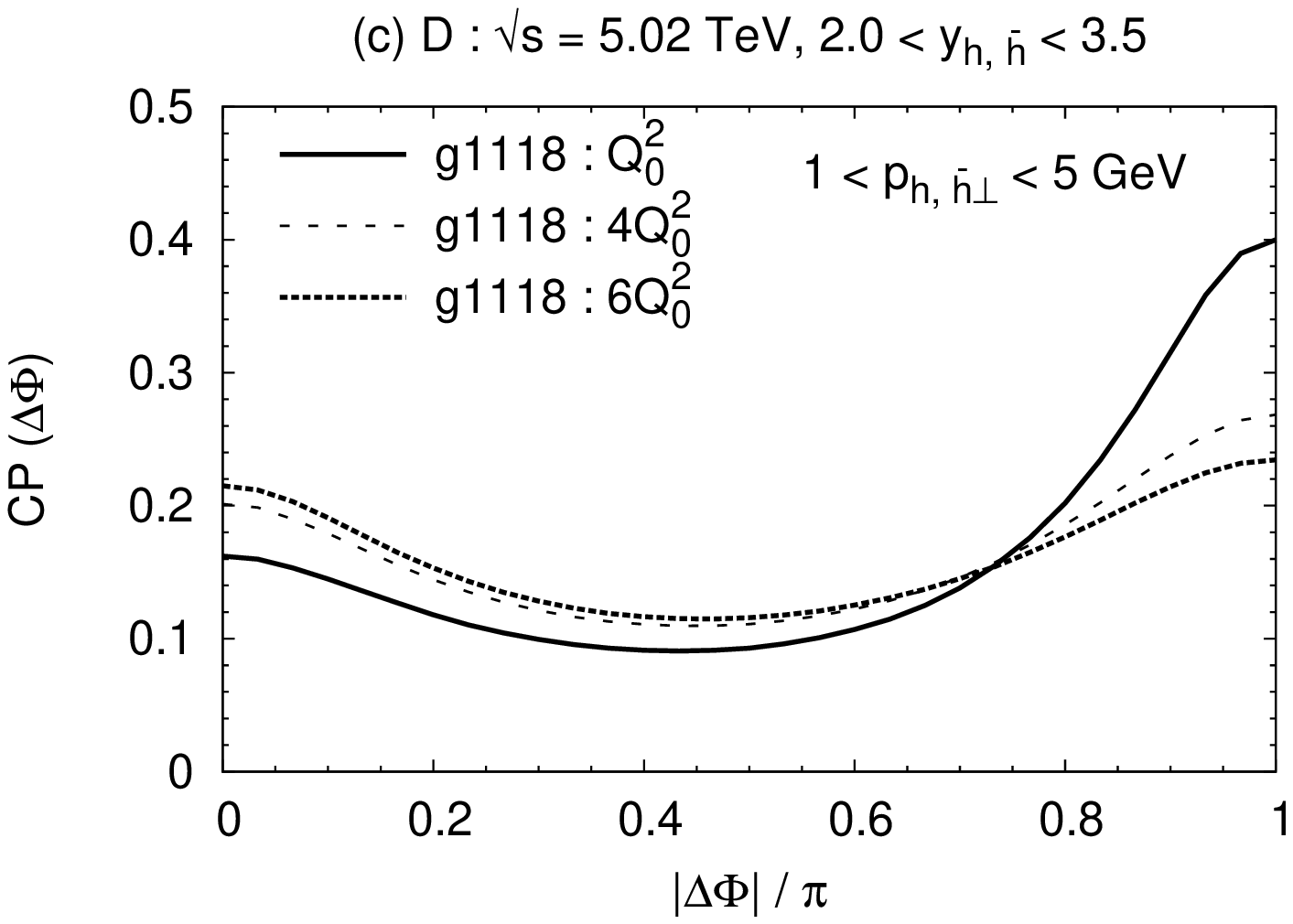}}
\resizebox*{!}{5.5cm}{\includegraphics[angle=-90]{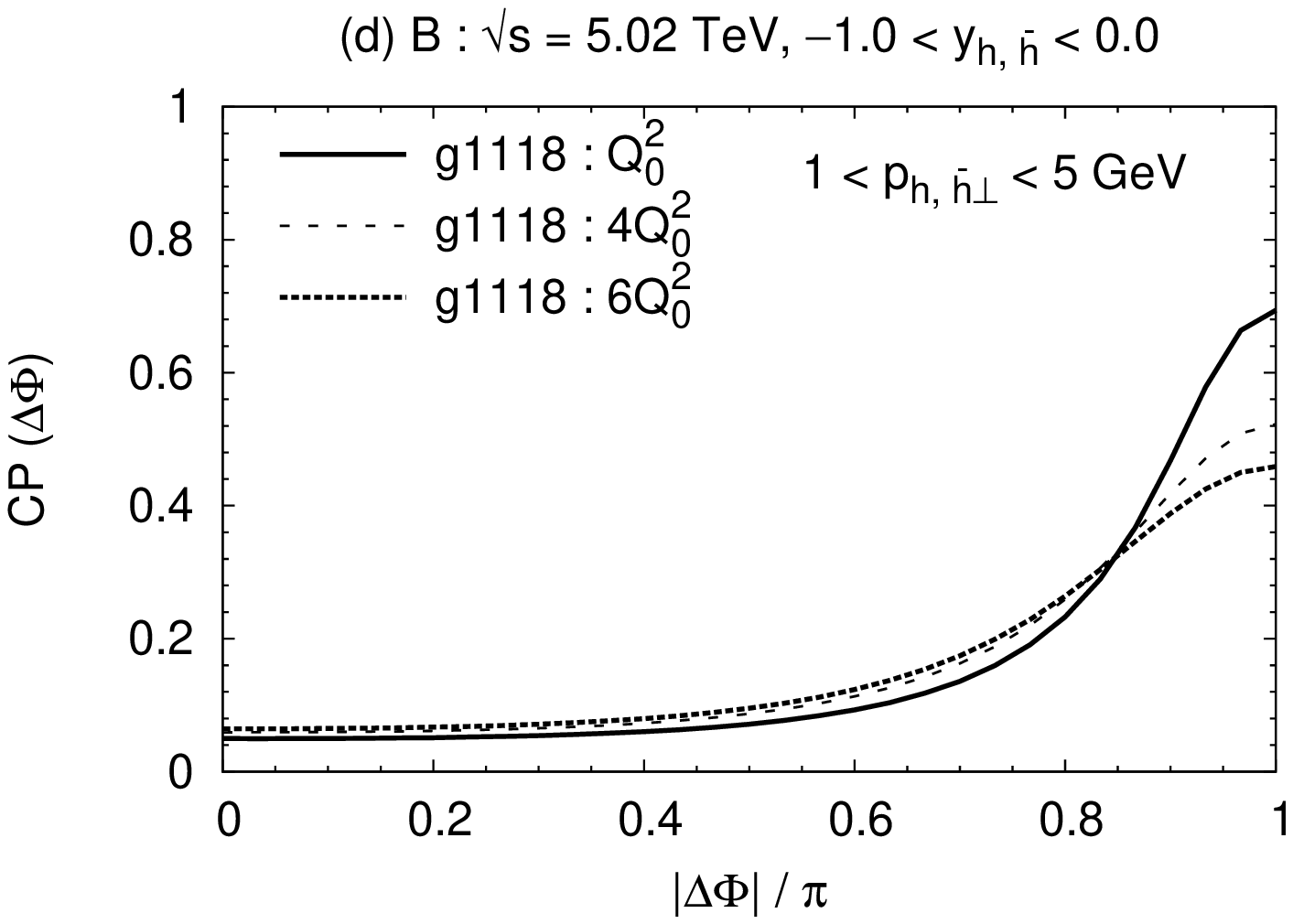}}
\end{center}
\caption{Nuclear modification of azimuthal angle correlation of heavy meson pair production 
in pA collision at $\sqrt s=5.02$ TeV.
Results with the initial saturation scale $Q_{0}^2$, $4 Q_0^2$ and 
$6 Q_0^2$ are plotted in solid, dashed and dotted lines, respectively.
(a) $D^0\bar D^0$ correlation with set g1118 for $-1<y_{h, \bar h}<0$,
(b) the same as (a) but with set MV,
(c) the same as (b) but for $2 < y_{h,\bar h} < 3.5$,
and
(d) $B^0\bar B^0$ correlation with set g1118 for $-1<y_{h,\bar h}<0$.
The momentum coverage is $1<p_{h, \bar h \perp}<5$ GeV, 
and $m_c=1.5$ GeV for $D^0 \bar D^0$ and
$m_b=4.8$ GeV for $B^0 \bar B^0$.
}
\label{fig:Correlation-s5020}
\end{figure}

Here we discuss
modification of the azimuthal angle correlation between open heavy flavor meson ($h$)
and open anti-flavor meson ($\bar h$) 
in pA collisions at $\sqrt s=5.02$ TeV. 
We set the momentum coverage to $1<p_{h\perp}, p_{\bar h\perp}<5$ GeV.

In Fig.~\ref{fig:Correlation-s5020}~(a)
we plot the numerical result obtained with uGD set g1118 
for the $D^0\bar D^0$ production at mid rapidity ($-1<y<0$).
The away-side peak seen at  $|\Delta\Phi|\sim\pi$ in pp collisions 
(initial scale $Q_0^2$)
is gradually suppressed in pA collisions 
with increasing the (initial) saturation scale in the nucleus
as $(4-6) Q_0^2$, while the near-side peak is slightly enhanced.
This is due to the stronger multiple scatterings and saturation effects in the heavy nucleus.
Then nuclear effects make $D^0\bar D^0$ correlation at low momentum
closer to isotropic distribution.
For comparison,
we show the same plot but with the uGD set MV in Fig.~\ref{fig:Correlation-s5020} (b).
Stronger enhancement of the correlation on the near side
than on the away side is seen with increasing the saturation scale of the uGD
in the nucleus. 
Different uGD sets result in quantitatively different correlation, but the qualitative
features remain the same.

The nuclear modification of the angle correlation becomes
more prominent 
in the forward rapidity region as seen in Fig. ~\ref{fig:Correlation-s5020} (c).
We have also computed the angle correlation in higher momentum region,
$5<p_{h,\bar h \perp}<10$ GeV. We saw a strong away-side peak suppressed in pA collisions
than in pp, while the near-side structure is unaffected.
Note that the transverse momentum on the near side is provided solely 
by the intrinsic $k_\perp$ of the gluons in (\ref{eq:open-cross-section-LN}).
The gluon saturation 
at  $k_\perp \lesssim Q_s$ does not affect the particle production
in such a high momentum region.

Finally, let us study 
$B^0\bar B^0$ correlations in the same kinematic region as $D^0 \bar D^0$, 
to see the quark mass dependence of the correlation.
As seen in Fig.~\ref{fig:Correlation-s5020} (d), 
despite that the momentum region is as low as 
in Fig.~\ref{fig:Correlation-s5020}~(a),
we do not confirm any correlation on the near side
since intrinsic momentum of gluon is still insufficient
to produce the pair there. 
The away-side peak exists and is suppressed with increasing $Q_{A,0}^2(x_0)$.

\section{Summary}

In this paper we have elaborated 
open heavy flavor meson production in high energy pA collisions
at RHIC and LHC energies in the Color Glass Condensate framework.
We have described the small-$x$ gluon distribution in the heavy nucleus
using the numerical solution of the rcBK equation which is constrained
with the HERA DIS data.

At RHIC energy, 
$D$ meson production proceeds from the moderate-$x$ gluons,
and there is not much room for $x$-evolution dynamics although
its forward-rapidity production has marginal 
sensitivity to $x<x_0=0.01$.
The nuclear modification factor $R_\text{pA}$ of $D$
shows a suppression at low $p_\perp$ and a Cronin-like enhancement
at larger $p_\perp$ reflecting the multiple scattering effects
implemented in the initial gluon distribution at $x=x_0$.
In contrast, 
$R_\text{pA}(p_\perp)$ of $B$ is almost flat in $p_\perp$.

At LHC energy, $D^0$ production with constrained gluon distribution
g1118 reasonably reproduces the $p_\perp$-dependence of
the mid-rapidity data in pp collisions.
The $R_\text{pA}$ of $D$ shows stronger suppression 
at low $p_\perp$ and no enhancement in computed $p_\perp$ region.
The $R_\text{pA}$ for $p_\perp$-integrated multiplicity
shows a systematic suppression in the forward rapidity region, 
which is due to the quantum $x$-evolution effect of the gluons in the
heavy nucleus.
We have also compared the $R_\text{pA}(y)$ of $D$ with that of
J/$\psi$ computed in the color evaporation model, and found that
J/$\psi$ production is more suppressed. This is because
the multiple scatterings additionally hinder the binding of the quark pair.

The $B^0$ production at the LHC is apparently overestimated by
a factor than the data at forward rapidity. 
We notice that the production involves the large-$x$ gluon distribution
in the proton, which we parametrize with a simple Ansatz.
Energy loss of the gluon in the nucleus may become also relevant there.
The $R_\text{pA}$ of $B$ still shows a suppression
with increasing rapidity $y$, but weaker than that of $D$.

As an unique candidate to study the gluon saturation in the nucleus, 
we have computed the azimuthal angle correlation 
for $D\bar D$ and $B\bar B$ pair in pp collision at LHC energy.
Because of the finite $k_{1,2\perp}$ of the incident gluons,
in addition to the away-side peak,
the near-side peak emerges in $D\bar D$ correlation, 
which is also seen in LHCb data\cite{double-charm}.
But we cannot quantitatively reproduce the angle correlation
and the invariant mass spectrum of the data at the same time.

In order to see the saturation effect on the angle correlation qualitatively,
we have calculated the $D^0\bar D^0$ correlation in pA collisions.
We have found that the away-side peak is more smeared 
and the near-side peak is slightly enhanced for the
larger saturation scale, i.e., with the heavier nucleus and/or
at more forward rapidity.
For $B \bar B$ correlation, we do not see the near-side peak because
the saturation scale is not large enough to produce 
the $B \bar B$ in the same azimuthal direction.



\section*{Acknowledgments}
The authors are very grateful to J.~Albacete, A.~Dumitru, F.~Gelis, 
K.~Itakura, Y.~Nara, R.~Venugopalan 
for useful discussions and collaborations on related topics.
They also thank members of Komaba Nuclear Theory Group for their
interests in this work.
This work was partially supported by 
Grant-in-Aids for Scientific Research ((C) 24540255) of MEXT.

\appendix

\section*{Appendix : Cutoff $z_\text{min}$}

The lower limit of the $z$ integration  in Eq.~(\ref{eq:open-cross-section-LN}) 
is given explicitly as
\begin{align}
 z_\text{min}
=\frac{q_{h\perp}\cosh y}{\sqrt{\frac{s}{4}-m^2\cosh^2 y}}
\ .
\end{align}
This can be readily derived 
by noting that the maximum energy of the produced quark and anti-quark 
in the center-of-mass frame is
$E_q^{\text{max}}=E_{\bar q}^{\text{max}}=\frac{\sqrt s}{2}$.
For the on-mass-shell quark, we have
\begin{align}
E_q^\text{max}&=\sqrt{m^2+(q_{\perp}^\text{max})^2}\cosh y
\ ,
\end{align}
where $y$ is the quark rapidity, which we set the same as the rapidity of the produced meson $h$.
Then $z_\text{min} \equiv q_{h\perp} / q_\perp^{\text{max}}$ gives the desired expression.


\end{document}